\documentclass[aps,prb,10pt,twocolumn,showpacs,showkeys,footinbib,superscriptaddress]{revtex4-2}

\usepackage[utf8]{inputenc}
\usepackage[T1]{fontenc}
\usepackage{amsmath}
\usepackage{amssymb}
\usepackage{graphicx}
\usepackage{placeins}
\usepackage[caption=false,position=top]{subfig}
\usepackage{pstricks}
\usepackage{color}
\usepackage{relsize}
\usepackage{bm}
\usepackage{braket}
\usepackage{epstopdf}
\usepackage{pgf}
\usepackage{hyperref}
\usepackage[all]{hypcap}

\renewcommand{\i}{\ensuremath{\mathrm{i}}}
\newcommand{\e}{\ensuremath{\mathrm{e}}}
\renewcommand{\d}{\ensuremath{\mathrm{d}}}

\newcommand{\sgn}{\operatorname{\mathrm{sgn}}}
\newcommand{\trace}{\operatorname{tr}}

\renewcommand{\vec}[1]{\bm{#1}}

\newcommand{\abs}[1]{\left|#1\right|}

\begin{document}
\title{Planar Josephson Hall effect in topological Josephson junctions}

\author{Oleksii Maistrenko}
\affiliation{Max-Planck-Institut f\"{u}r Festk\"{o}rperforschung, D-70569 Stuttgart, Germany}
\author{Benedikt Scharf}
\affiliation{Institute for Theoretical Physics and Astrophysics and W\"{u}rzburg-Dresden Cluster of Excellence ct.qmat, University of W\"{u}rzburg, Am Hubland, 97074 W\"{u}rzburg, Germany}
\author{Dirk Manske}
\affiliation{Max-Planck-Institut f\"{u}r Festk\"{o}rperforschung, D-70569 Stuttgart, Germany}
\author{Ewelina M.~Hankiewicz}
\affiliation{Institute for Theoretical Physics and Astrophysics and W\"{u}rzburg-Dresden Cluster of Excellence ct.qmat, University of W\"{u}rzburg, Am Hubland, 97074 W\"{u}rzburg, Germany}

\date{\today}

\begin{abstract}
  Josephson junctions based on three-dimensional topological insulators offer intriguing possibilities to realize unconventional $p$-wave pairing and Majorana modes.
  Here, we provide a detailed study of the effect of a uniform magnetization in the normal region: We show how the interplay between the spin-momentum locking of the topological insulator and an in-plane magnetization parallel to the direction of phase bias leads to an asymmetry of the Andreev spectrum with respect to transverse momenta.
  If sufficiently large, this asymmetry induces a transition from a regime of gapless, counterpropagating Majorana modes to a regime with unprotected modes that are unidirectional at small transverse momenta.
  Intriguingly, the magnetization-induced asymmetry of the Andreev spectrum also gives rise to a Josephson Hall effect, that is, the appearance of a transverse Josephson current.
  The amplitude and current phase relation of the Josephson Hall current are studied in detail.
  In particular, we show how magnetic control and gating of the normal region can enable sizable Josephson Hall currents compared to the longitudinal Josephson current.
  Finally, we also propose in-plane magnetic fields as an alternative to the magnetization in the normal region and discuss how the planar Josephson Hall effect could be observed in experiments.
\end{abstract}

%\pacs{...}
%\keywords{...}

\maketitle

\section{Introduction}\label{Sec:Intro}
The helical spin structure of the surface states of three-dimensional topological insulators (3D TIs) offers intriguing possibilities of tailoring the surface-state properties by various proximity effects.
A conventional $s$-wave superconductor can, for example, be used to proximity-induce superconductivity in the TI surface.
The interplay between the helical spin-momentum locking of the TI surface state and the superconducting pairing then mediates an effective pairing between electrons at the Fermi level.
This effective pairing features a mixture of singlet $s$-wave and triplet $p$-wave pair correlations~\cite{Fu2008:PRL,Alicea2012:RPP,Tkachov2013:PRB} and turns the TI surface into a topological superconductor~\cite{Hasan2010:RMP,*Qi2011:RMP,Alicea2012:RPP,Leijnse2012:SST,Tanaka2012:JPSJ,Beenakker2013:ARCMP,Tkachov2013:PSS,Culcer2020:2DM} with Majorana zero modes~\cite{Fu2008:PRL} and odd-frequency pairing~\cite{BlackSchaffer2012:PRB}.

In this context, Josephson junctions based on 3D TIs or on their two-dimensional (2D) counterparts have been studied extensively for potential signatures of topological superconductivity, both theoretically~\cite{Fu2008:PRL,Tanaka2009:PRL,Houzet2013:PRL,Beenakker2013:PRL,Tkachov2013:PRB,Crepin2014:PRL,Tkachov2015:PRB,Sothmann2016:PRB,Tkachov2017:PRB,Tkachov2019:PRB,*Tkachov2019:JPCM,PicoCortes2017:PRB,Dominguez2017:PRB,Murani2019:PRL,Zhang2020:PRR,Keidel2019:arxiv,Calzona2019:PRR} and experimentally~\cite{Wiedenmann2016:NC,Kayyalha2019:PRL,Oostinga2013:PRX,Sochnikov2015:PRL,Deacon2017:PRX}.
These so-called topological Josephson junctions exhibit a ground-state fermion parity that is $4\pi$-periodic in the superconducting phase difference $\phi$ and Andreev bound states (ABS) with a protected zero-energy crossing~\cite{FuKane2009:PRB,Ioselevich2011:PRL}.

\begin{figure}[ht]
  \centering
  \includegraphics[width=\columnwidth]{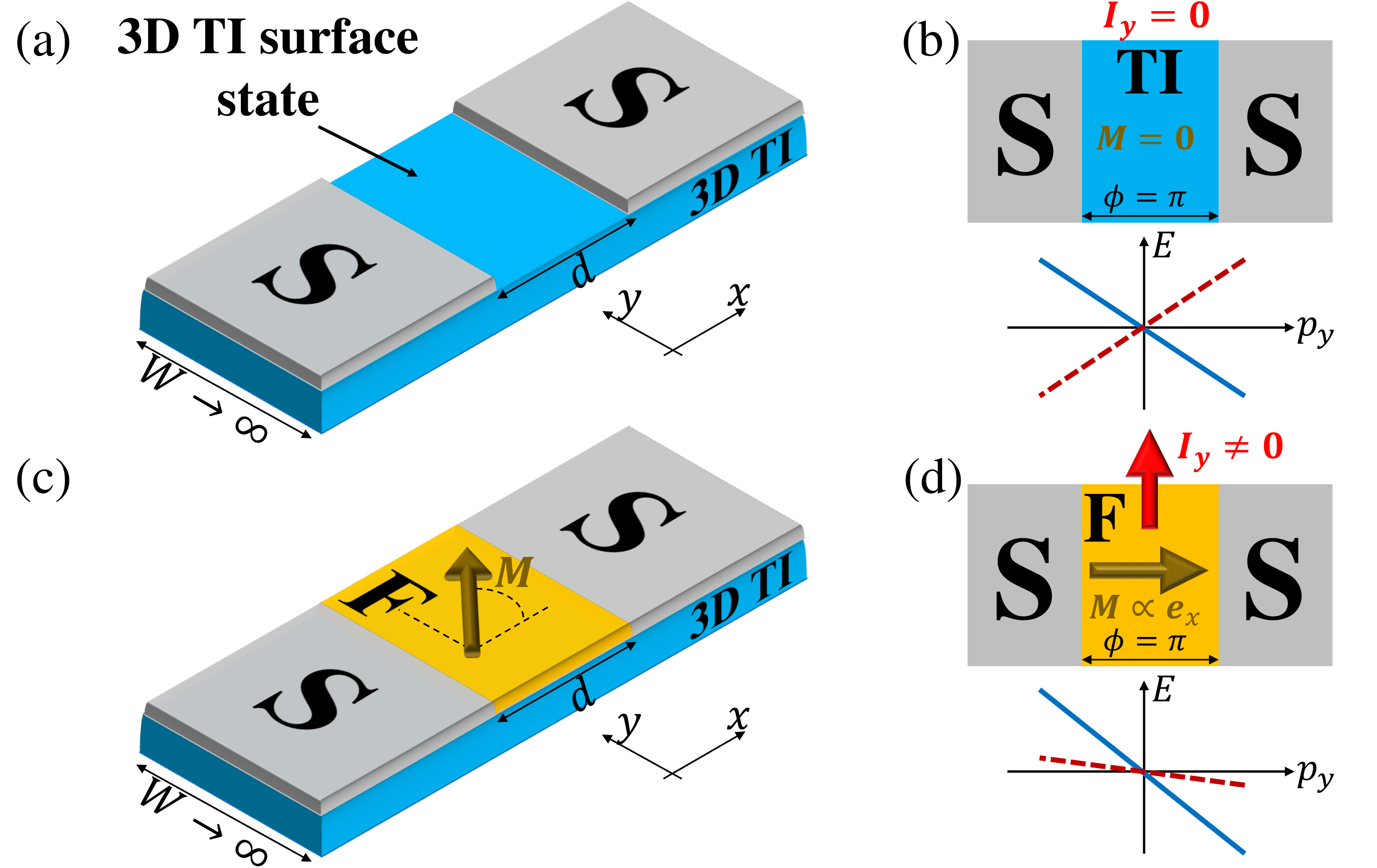}
  \caption{(Color online) (a) Scheme of a Josephson junction based on a 3D TI: $s$-wave superconductors (S) on top of the TI proximity-induce pairing into the TI surface state.
    The two proximity-induced superconducting regions are separated by a normal region.
    (b) Top view and low-energy Andreev spectrum of a short topological $\pi$-junction for transverse momenta $p_y$ close to $p_y=0$: The two low-energy ABS correspond to counterpropagating nonchiral Majorana modes with opposite group velocities.
    No net Josephson Hall current flows in $y$ direction.
    (c) Ferromagnetic Josephson junction based on a 3D TI: Same as (a), but with a magnetic region separating the superconducting regions.
    In this setup, the Zeeman field/exchange splitting $\bm{M}$ is proximity-induced by a ferromagnet (F).
    (d) Same as (a), but for a ferromagnetic Josephson junction with large $\bm{M}$ parallel to the direction of the phase bias: The Andreev spectrum is asymmetric and has been tilted in such a way that the two low-energy modes are unidirectional for small $p_y$.
    Note that these ABS are no Majorana modes protected against backscattering because there are additional zero-energy states for $p_y$ close to the Fermi momentum (not shown).
    The asymmetry in the Andreev spectrum gives rise to a finite Josephson Hall current flowing in $y$ direction.
  }\label{fig:scheme}
\end{figure}

Topological Josephson junctions are particularly intriguing if they are based on 3D TIs, as depicted in Fig.~\ref{fig:scheme}(a): Because of the 2D nature of the surface, the system supports modes that propagate along the direction parallel to the superconductor/normal TI interface, that is, the $y$ direction in Fig.~\ref{fig:scheme}(a).
Due to the protected zero-energy crossing occurring at zero transverse momentum and phase difference $\phi=\pi$, a $\pi$-junction exhibits two counterpropagating, gapless states, so-called nonchiral Majorana modes~\cite{Fu2008:PRL} [see Fig.~\ref{fig:scheme}(b) bottom].

Besides proximity-induced superconductivity, one can also envision other proximity effects whose interplay with the spin texture of the TI surface state leads to novel phenomena: In non-superconducting setups, for example, the interplay between the helical surface states and proximity-induced magnetism provides a versatile platform for studying fundamental effects and spintronic applications~\cite{Hasan2010:RMP,*Qi2011:RMP,Shen2012,Mellnik2014:N,*Fan2014:NM}.
Ferromagnetic tunnel junctions based on 3D TIs~\cite{Mondal2010:PRL,Wu2010:PRB3,*Wu2012:NRL,Li2014:PRB,*Li2014:NN,Tian2014:SSC,*Tian2015:NSR,Scharf2016:PRL}, in particular, show some promise for potential spintronic devices~\cite{Zutic2004:RMP,*Fabian2007:APS}.
The combination of 3D TIs with both proximity-induced superconductivity and magnetism can prove even more interesting~\cite{Tanaka2009:PRL,Linder2010:PRB,Snelder2013:PRB,Burset2015:PRB}, however, and could point to novel possibilities for superconducting spintronics~\cite{Eschrig2011:PT,Linder2015:NP}.

%\cite{Kastening2006:PRL,Brydon2008:PRB,Brydon2009:PRL}.
Motivated by this prospect~\footnote{The interplay between triplet pairing and ferromagnetism is, moreover, known to give rise to novel types of Josephson effect and Josephson current switches [B. Kastening, D. K. Morr, D. Manske, and K. Bennemann, Phys. Rev. Lett. 96, 047009 (2006); P. M. R. Brydon, B. Kastening, D. K. Morr, and D. Manske, Phys. Rev. B 77, 104504 (2008); P. M. R. Brydon and D. Manske, Phys. Rev. Lett. 103, 147001 (2009)].} as well as by phenomena found in non-superconducting TI tunneling junctions, such as the tunneling planar Hall effect~\cite{Scharf2016:PRL}, we study 3D TI-based Josephson junctions with a ferromagnetic tunneling barrier [see Fig.~\ref{fig:scheme}(c)].
In contrast to previous studies on this system~\cite{Tanaka2009:PRL,Linder2010:PRB,Snelder2013:PRB}, we focus not only on the longitudinal response, but also on the transverse response to an applied phase bias.
We find that especially the configuration with an in-plane magnetization parallel to the direction of the phase bias exhibits striking features: Such a magnetization leads to an asymmetric Andreev spectrum for a fixed finite transverse momentum.
If sufficiently large, this asymmetry even induces a transition from the regime of counterpropagating, nonchiral Majorana modes to a regime with unprotected unidirectional modes at small transverse momenta [compare Fig.~\ref{fig:scheme}(b) bottom and Fig.~\ref{fig:scheme}(d) bottom].
Most importantly, even a small magnetization-induced asymmetry in the Andreev spectrum causes a transverse Josephson Hall current [see Fig.~\ref{fig:scheme}(d) top].
In contrast to other Josephson Hall effects~\cite{Yokoyama2015:PRB,Malshukov2019:PRB}, the effect found here arises from an in-plane magnetization, which is why we call it the planar Josephson Hall effect.
The planar Josephson Hall effect is the superconducting analog to the tunneling planar Hall effect found in non-superconducting TI tunneling junctions~\cite{Scharf2016:PRL}.

Below, we will discuss the origin of the Josephson Hall current, its properties and how it could be experimentally verified.
The manuscript is organized as follows: After introducing the effective model used to describe the Josephson junction in Sec.~\ref{Sec:Model}, we study its ABS in Sec.~\ref{Sec:ABS}.
In Secs.~\ref{Sec:CO} and~\ref{Sec:GF}, the procedure to compute the different Josephson currents is presented.
These currents are then discussed in Sec.~\ref{Sec:CPR}.
A brief summary concludes the manuscript in Sec.~\ref{Sec:Conclusions}.

\section{Model}\label{Sec:Model}
\subsection{Hamiltonian and unitary transformation}
In our model, we consider a Josephson junction based on the 2D surface state of a 3D TI, as depicted in Fig.~\ref{fig:scheme}(c), where the pairing in the superconducting (S) regions is induced from a nearby $s$-wave superconductor.
The ferromagnetic (F) region is subject to an exchange splitting/Zeeman term proximity-induced from a nearby ferromagnet~\cite{Zutic2019:MT}.
If one is only interested in an in-plane Zeeman term, an alternative way to realize such a Zeeman term is by applying an in-plane magnetic field as discussed in Sec.~\ref{Sec:Conclusions} below.
The surface state lies in the $xy$ plane, with the direction of the superconducting phase bias denoted as the $x$ direction.
We take the system to be infinite in both the $x$ and $y$ directions.
Here, we study the regime where the Fermi level is situated inside the bulk gap and where only surface states exist.
Moreover, we assume that the surface considered is far enough away from the opposite surface so that there is no overlap between their states.
Then, the Josephson junction based on a single surface is described by the Bogoliubov-de Gennes (BdG) Hamiltonian
\begin{multline}\label{eq:HBdGHam}
  \hat{H}_\mathrm{BdG}^0=\left[v_F\left(\sigma_x\hat{p}_y-\sigma_y\hat{p}_x\right)-\mu\right]\tau_z+(V_0\tau_z-\bm{M}\cdot\bm{\sigma})h(x)\\
  +\Delta(x)\left[\tau_x\cos\Phi(x)-\tau_y\sin\Phi(x)\right]
\end{multline}
with the basis order $\hat\Psi=\left(\hat{\psi}_\uparrow,\hat{\psi}_\downarrow,\hat{\psi}^\dagger_\downarrow,-\hat{\psi}^\dagger_\uparrow\right)^T$.
In Eq.~(\ref{eq:HBdGHam}), $\hat{p}_l$ (with $l=x,y$) denote momentum operators and $\sigma_l$ and $\tau_l$ (with $l=x,y,z$) Pauli matrices in spin and particle-hole space, respectively.
Moreover, $\bm{\sigma}=(\sigma_x,\sigma_y,\sigma_z)$ and unit matrices are not written explicitly in Eq.~(\ref{eq:HBdGHam}).

In this manuscript, we study two models for a Josephson junction with a F region of width $d$: (a) a model with a $\delta$-like F region described by $h(x)=d\delta(x)$ and $\Delta(x)=\Delta$ and (b) a model with a finite F region where $h(x)=\Theta(d/2-|x|)$ and $\Delta(x)=\Delta\Theta(|x|-d/2)$.
In both cases, the phase convention is $\Phi(x)=\phi\Theta(x)$, where $\phi$ is the superconducting phase difference between the two S regions.
The superconducting pairing amplitude with strength $\Delta\geq0$ is proximity-induced from the $s$-wave superconductors deposited on the TI surface.
The density of states of these superconductors is typically much larger than that of the TI surface states, e.g we can assume Nb superconductors used in experiments.
Therefore, the currents flowing in the TI surface do not significantly affect the superconducting phases 
and we can use constant $\phi$ and $\Delta$ within each superconducting lead.
In other words the resistivity of junction region is much larger than that of the leads,
this justifies our approximation~\cite{Likharev1979:RMP,Beenakker1997:RMP} commonly used in mesoscopic systems.
We note that the superconductors from which superconductivity is induced in the TI surface states are not explicitly included in our model~(\ref{eq:HBdGHam}).
However, in the real system they are important to make this assumption.
The Fermi velocity of the surface state is $v_F$, and $V_0$ the potential in the F region, which can also be viewed as describing the difference between the chemical potentials in the S and F regions, $\mu$ and $\mu_\mathrm{F}=\mu-V_0$.
The Zeeman term due to the proximity-induced ferromagnetic exchange splitting is described by the effective magnetization $\bm{M}=(M_x,M_y,M_z)$~\cite{Zutic2019:MT}.
Note that the direction of $\bm{M}$ is set by the magnetization in the ferromagnet.

For our calculations, it is more convenient to introduce the unitary rotation transformation in spin space $U=(1-\i\sigma_z)/\sqrt{2}$ and bring the Dirac Hamiltonian~(\ref{eq:HBdGHam}) into the form
\begin{multline}\label{eq:HBdGHamRot}
  \hat{H}_\mathrm{BdG}=\left[
    \vec\sigma\cdot \left(v_F \hat{\vec p} - \tau_z \vec{M}' h(x) \right)
    -\mu + V_0 h(x)
    \right]\tau_z \\
  +\Delta(x)\left[\tau_x\cos\Phi(x)-\tau_y\sin\Phi(x)\right]
\end{multline}
with $\hat{\vec p}=(\hat{p}_x,\hat{p}_y,0)$.
Because of the rotated spin axes used in Eq.~(\ref{eq:HBdGHamRot}) $\bm{M}'$ is a rotated effective magnetization,
which is related to the components of the real magnetization $\bm{M}$ induced in the F region via $\bm{M}'=(-M_y,M_x,M_z)$.
In addition, for the Dirac equation this Zeeman term has the same form as a vector potential.
From now on, we use the Hamiltonian~(\ref{eq:HBdGHamRot}) because it proves more convenient mathematically.

\subsection{General form of the solutions}
To solve $\hat{H}_\mathrm{BdG}\Psi(\bm{r})=E\Psi(\bm{r})$ and obtain the eigenspectrum of Eq.~(\ref{eq:HBdGHamRot}), we first make use of translational invariance along the $y$ direction, $[\hat{H}_\mathrm{BdG},\hat{p}_y]=0$.
Although, on a macroscopic scale $y \gg d$ the phase may depend on the transverse coordinate,
this should not affect the local structure of Andreev levels calculated below.
Hence, we proceed with the ansatz $\Psi(\bm{r})=\e^{\i p_yy}\psi(x)/\sqrt{W}$,
which reasonably simplifies the analytical treatment of the system.
Here $p_y$ is the momentum quantum number, $\psi(x)$ is a spinor in Nambu space, and $W$ is a unit width of the system in $y$ direction.
%This ansatz allows us to calculate the junction properties far away from the top and bottom edges of the system.
Even if one considers a large finite-size system in the $y$ direction, these solutions should describe states away from the boundaries.
Here and in the remainder of this manuscript, we set $\hbar=1$.
The eigenenergies and $\psi(x)$ can then be obtained from the 1D BdG equation
\begin{equation}\label{eq:BdG}
\hat{H}_\mathrm{BdG}(p_y)\psi(x)=E\psi(x),
\end{equation}
where $\hat{H}_\mathrm{BdG}(p_y)$ is given by Eq.~(\ref{eq:HBdGHamRot}) with the operator $\hat{p}_y$ replaced by the quantum number $p_y$.

The energy-momentum relation in the S regions is given by $q_{\pm}=\sqrt{(\mu\pm\Omega)^2/v_F^2-p_y^2}$ with $\Omega=\sqrt{E^2-\Delta^2}$.
We find the following solutions in the S leads:
\begin{equation}\label{eq:PsiLead}
\psi^{(S)}_{\xi\alpha}(x)=\frac{1}{\sqrt{2}}\begin{pmatrix}u_{\xi} \\
  e^{-i\Phi(x)}v_{\xi}
\end{pmatrix}\text{\ensuremath{\otimes}}\begin{pmatrix}1 \\
  \dfrac{v_F(\alpha q_{\xi}+ip_{y})}{\mu+\xi\Omega}
\end{pmatrix}e^{i\alpha q_{\xi}x},
\end{equation}
where $\xi=\pm1$ corresponds to particle-like and hole-like solutions and $\alpha=\pm1$ selects the direction of motion.
Here,
\begin{equation}
u_{\xi} =\sqrt{\frac{1}{2}\left(1+\frac{\xi\Omega}{E}\right)},\quad
v_{\xi} =\sqrt{\frac{1}{2}\left(1-\frac{\xi\Omega}{E}\right)}.
\end{equation}

In the F region, the electron and hole states are given by
\begin{equation}\label{eq:PsiNormal}
\psi^{(F)}_{\xi\alpha}(x)=\frac{e^{i(-\xi M_{y}/v_F+\alpha k_{\xi})x}}{\sqrt{2E'(E'-\xi M_{z})}}
\begin{pmatrix}E'-\xi M_{z} \\
  \alpha v_Fk_{\xi}+i(v_Fp_{y}-\xi M_{x})
\end{pmatrix}
\end{equation}
\\ % small space
with $v_Fk_{e/h}=\sqrt{(\mu\pm E-V_0)^2-(v_Fp_y\mp M_x)^2-M_z^2}$ and $E'=\mu+\xi E-V_0$.
The Zeeman term $M_z$ opens a symmetric gap in the spectrum,
while in-plane magnetization $M_x$ shifts the position of the Dirac cone and introduces an asymmetry in the barrier states.
For a given $p_y$ mode this changes the effective energy gap in the barrier,
making the Andreev reflection process angle dependent.
%In other words, component of the effective magnetization $\vec M'$ perpendicular to the propagation direction act as mass terms.
This result of spin-momentum locking will have important consequences for the discussion below.

\section{Andreev bound states}\label{Sec:ABS}
\subsection{General equations}\label{Sec:ABSGE}
In order to understand the Josephson currents and the emergence of a Josephson Hall current, it is instructive to first look at the ABS of Eq.~(\ref{eq:BdG}), that is, bound states decaying for $|x|\to\infty$ and hence with energies $|E|<\Delta$.
We focus on the ABS of a junction with finite F region and refer to Appendix~\ref{App:ABSdelta} for the Andreev spectrum of the $\delta$-model, where relatively compact, analytical solutions are possible in certain limiting cases.
The eigenenergies of the ABS and their corresponding eigenstates can be determined from the ansatz
\begin{equation}\label{eq:ABSansatz}
\psi(x)=\begin{cases}
  A_1 \psi^{(S)}_{e,-s_\mu}(x) + A_2 \psi^{(S)}_{h,s_\mu}(x), \;        & x < -\frac{d}{2}      \\[5pt]
  \sum_{\xi=e/h,\alpha=\pm} D_{\xi\alpha} \psi^{(F)}_{\xi\alpha}(x), \; & \abs{x} < \frac{d}{2} \\[5pt]
  B_1 \psi^{(S)}_{e,s_\mu}(x) + B_2 \psi^{(S)}_{h,-s_\mu}(x), \;        & \frac{d}{2} < x
\end{cases}
\end{equation}
for a junction with a finite F region and $s_\mu=\sgn(\mu)$.
Now, the coefficients $A_1$, $A_2$, $D_{e\pm}$, $D_{h\pm}$, $B_1$, $B_2$ have to be calculated from the boundary conditions at the S/F interfaces,
\begin{equation}\label{eq:BCfinite}
\psi(0^+)=\psi(0^-),\quad\psi(d^+)=\psi(d^-).
\end{equation}
The boundary conditions~(\ref{eq:BCfinite}) lead to systems of linear equations for the coefficients $A_1$ to $B_2$.
By requiring a nontrivial solution of this system of linear equations, that is, by requiring its determinant to vanish, we find the ABS energies $E=E(\phi,p_y)$.

\subsection{Andreev spectrum of a ferromagnetic Josephson junction}\label{Sec:ABSnum}
\begin{figure}[t]
  \centering
  \includegraphics[]{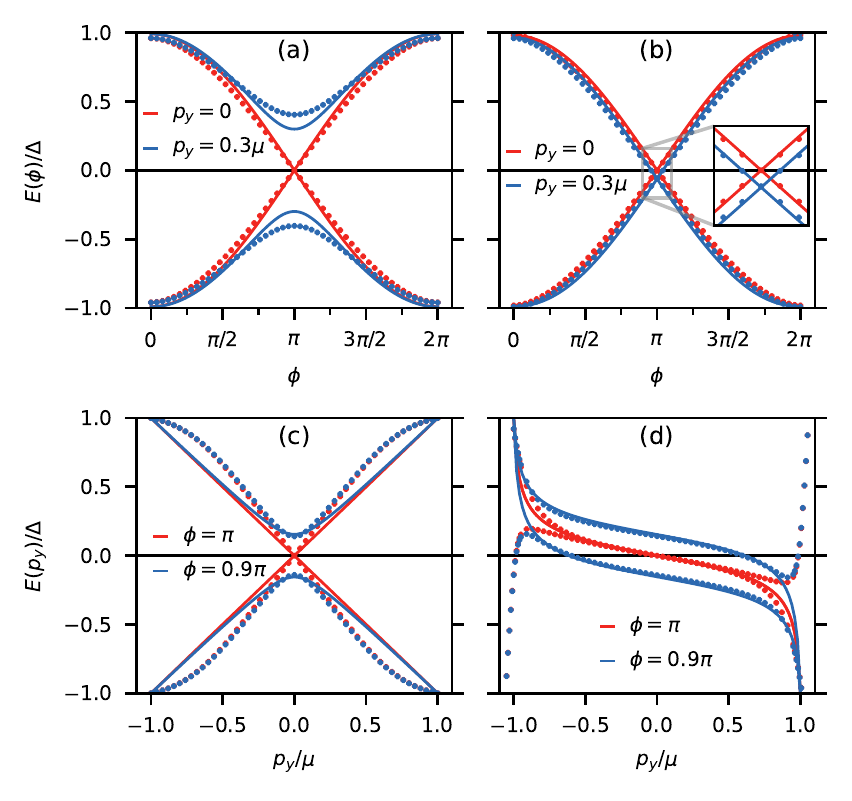}
  \caption{(Color online) Andreev bound state spectra for different combinations of $\bm{M}$ and $V_0$: (a,c) $\bm{M}=M\bm{e}_z$, $V_0=1.5$ meV and (b,d) $\bm{M}=M\bm{e}_x$, $V_0=0$.
    Here, $\bm{e}_l$ denotes a unit vector in $l$ direction with $l=x,y,z$.
    In all panels, $M=0.2$ meV, $d=330$ nm, $\mu=2$ meV, $v_F=5\times10^5$ m/s, and $\Delta=100$ $\mu$eV.
    The solid lines depict the spectra given in Eq.~(\ref{eq:SolAA}) for the $\delta$-barrier in the Andreev approximation.
    The discrete data points depict the numerically computed ABS spectra as obtained for the finite F region without any approximations to Eq.~(\ref{eq:HBdGHamRot}).
  }\label{fig:ABS}
\end{figure}

This procedure enables us to compute the Andreev spectrum of a finite barrier, examples of which are shown in Fig.~\ref{fig:ABS} for a short junction with a F region of length $d=330$ nm, $|\mu|\gg\Delta$, and different configurations of $\bm{M}$.
For these parameters, there are two ABS with energies $E_{\pm}(\phi,p_y)$ at a given momentum $p_y$, where the subscript $\pm$ denotes which state lies higher (lower) in energy, that is, $E_+(\phi,p_y)\geq E_-(\phi,p_y)$.
We can compare the $\phi$ and $p_y$ dependence of these ABS with the case of no magnetization, that is, $\bm{M}=0$ (not shown): For $\bm{M}=\bm{0}$, the Andreev spectrum $E_\pm(\phi,p_y)$ exhibits a zero-energy crossing protected by fermion parity at odd integer multiples of $\phi=\pi$ and $p_y=0$, as also discussed in Appendix~\ref{App:ZEC}.
This protected zero-energy crossing is accompanied by two gapless, nonchiral Majorana modes that counterpropagate along the $y$ direction and are localized mostly in the normal region~\cite{Fu2008:PRL,Tkachov2013:PRB}.

If we include a finite $\bm{M}$, its effects on the ABS are the following:

i) A component $M_y$ (not shown) shifts the entire Andreev spectrum as a function of $\phi$, that is, $E_{\pm}(\phi,p_y)\to E_{\pm}(\phi + 2Z_y,p_y)$, where $Z_y=M_y d/v_F$, but leaves the spectrum otherwise unchanged \cite{Tanaka2009:PRL}.
In particular, the protected zero-energy crossing for $p_y=0$ and the nonchiral Majorana modes are now shifted to $\phi=(2n+1)\pi-2Z_y$, where $n\in\mathbb{Z}$ is an integer.
Indeed, the $M_y$ component can be absorbed into the phase difference by performing a gauge transformation of the BdG Hamiltonian (see Appendix \ref{App:ABSdelta}).

ii) Finite components $M_x$ and $M_z$, shown in Figs.~\ref{fig:ABS}(b,d) and~(a,c) respectively, also do not remove this zero-energy crossing for $p_y=0$ and $\phi=(2n+1)\pi-2Z_y$.
This crossing remains protected by the fermion parity and cannot be removed by a finite $M_x$ or $M_z$~\cite{FuKane2009:PRB,Ioselevich2011:PRL} (see also Appendix~\ref{App:ZEC}).
The main effect of a finite out-of-plane magnetization $M_z$ in the F region is to detach the ABS from the continuum states with $|E|>\Delta$ [see Fig.~\ref{fig:ABS}(a) and Appendix \ref{App:ABSlargeMz}], consistent with the results found in Refs.~\cite{Linder2010:PRB,Snelder2013:PRB}.

iii) Intriguingly, we find that a finite $M_x\neq0$ introduces an asymmetry in the Andreev spectrum at finite $p_y$ as shown in Figs.~\ref{fig:ABS}(c) and~(d): It does no longer satisfy $E_{\pm}(\phi,p_y)=-E_{\mp}(\phi,p_y)$, but only the weaker condition $E_{\pm}(\phi,p_y)=-E_{\mp}(\phi,-p_y)$, dictated by the particle-hole symmetry of the BdG formalism.
In particular, Fig.~\ref{fig:ABS}(d), which shows the $p_y$ dependence of the Andreev spectrum, illustrates that the asymmetry $E_{\pm}(\phi,p_y)\neq-E_{\mp}(\phi,p_y)$ manifests itself in a 'tilting' of the spectrum.
If $M_x$ is large enough, it can even lead to a situation where the group velocities in $y$ direction, $v_\mathrm{g}\propto\partial E_{\pm}(\phi,p_y)/\partial p_y$, for ABS in the vicinity of $p_y=0$ and $\phi\approx\pi-2Z_y$ have the same sign.
Such a situation is shown in Fig.~\ref{fig:ABS}(d).
In this regime, the ABS change from nonchiral, counterpropagating Majorana modes to modes propagating in the same direction for small $p_y$.
At small $p_y$, the dispersion of these ABS is reminiscent of the unidirectional modes found in noncentrosymmetric superconductors~\cite{Wong2013:PRB,Daido2016:PRB,Daido2017:PRB} or in Rashba sandwiches~\cite{Volpez2018:PRB}.
An energy spectrum asymmetric in the transverse momentum $p_y$ can also appear at a single F/S interface due to broken rotational symmetry by the $M_x$ term \cite{Burset2015:PRB}.
It is important to note that the unidirectional ABS close to $p_y=0$ are, however, not protected against backscattering: As can be seen in Fig.~\ref{fig:ABS}(d), these states are accompanied by other zero-energy states with $p_y$ close to the Fermi momentum and with opposite group velocities.

The results presented above show that although the Zeeman term enters the equations in the form of a vector potential [see Eq.~\ref{eq:HBdGHamRot}],
its effect is not limited to the semi-classical phase factor typical for a spin degenerate electron system.
The found asymmetry of the Andreev spectrum emerges from the interplay between the spin-orbit coupling of the TI and $M_x$ which plays the role of the magnetic tunneling barrier.
We discuss it in more detail in Appendix~\ref{App:ABSeffM} with an effective low-energy model.

In Fig.~\ref{fig:ABS}, we also compare the numerically obtained ABS with the analytical expressions one can derive for the ABS of a model with a $\delta$-like F region in the Andreev approximation, as discussed in Appendix~\ref{App:ABSdelta}.
For short junctions and momenta close to $p_y=0$, these analytical expressions provide an excellent description of the ABS.
In particular, these expressions also capture the asymmetry and 'tilting' of the Andreev spectrum induced by $M_x$.
In addition, we show in Appendix~\ref{App:ABSsmallmu} that the same effect is also present for parameters beyond the Andreev approximation, i.e for $\mu\sim\Delta$.

\subsection{Spin structure of Andreev bound states}

\begin{figure}[t]
  \centering
  \includegraphics[width=\columnwidth]{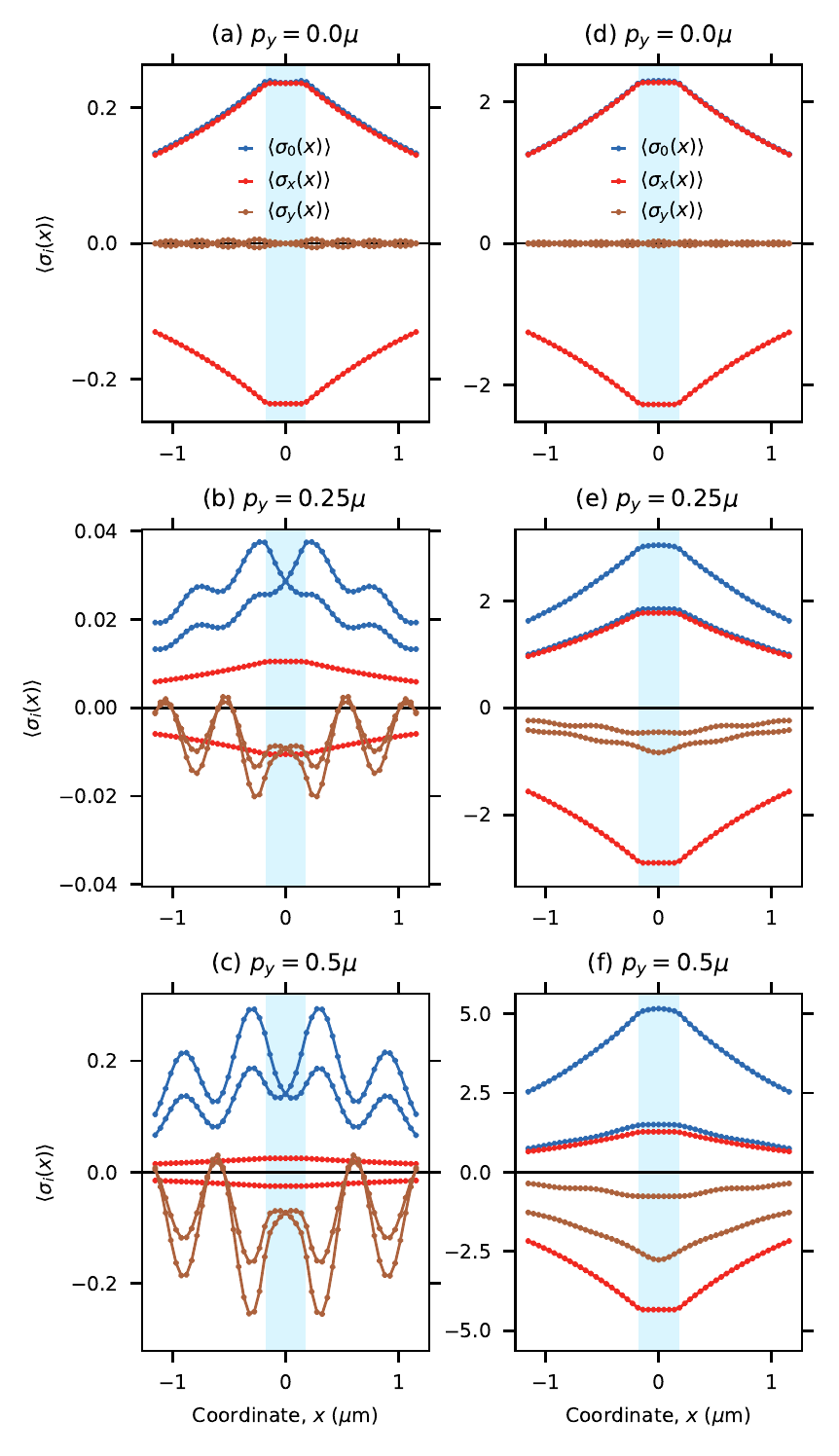}
  \caption{(Color online) Expectation values $\sigma_i(x)$ obtained from the ABS wave functions: (a-c) $\bm{M}=M\bm{e}_z$, $V_0=1.5$ meV, (d-f) $\bm{M}=M\bm{e}_x$, $V_0=0$.
    Here, $\bm{e}_l$ denotes a unit vector in $l$ direction with $l=x,y,z$.
    In all panels, $\phi=0.9\,\pi$, $M=0.2$ meV, $d=330$ nm, $\mu=2$ meV, $v_F=5\times10^5$ m/s, and $\Delta=100$ $\mu$eV.
    Light blue designates the normal (F) region.
  }\label{fig:ABSx}
\end{figure}

Figure~\ref{fig:ABSx} shows the spatial dependence of the quasiparticle density $|\psi(x)|^2=\langle\sigma_0(x)\rangle$ of the two ABS for $\phi=0.9\pi$ and different momenta $p_y$ if $M_z\neq0$ [Figs.~\ref{fig:ABSx}(a-c)] and if $M_x\neq0$ [Figs.~\ref{fig:ABSx}(d-f)].
As can be discerned from Figs.~\ref{fig:ABSx}(a-c), $M_z\neq0$ leads to ABS that are increasingly localized at the S/F interfaces as $p_y$ or $M_z$ are increased.
One can understand this behavior by recalling that a magnetization component in $z$ direction acts as a mass term that increasingly isolates the left and right S regions.
If the two S regions are completely isolated from each other, that is, for $M_z\to\infty$, each S region separately corresponds to a topological superconductor that hosts one chiral Majorana mode at its boundary~\cite{Fu2008:PRL}.
Hence, the results in Figs.~\ref{fig:ABSx}(a-c) can be interpreted as the intermediate regime between $\bm{M}=\bm{0}$ with nonchiral Majorana modes that are completely delocalized inside the F region and $M_z\to\infty$ with one chiral Majorana mode at each of the S/F interfaces.
Comparing $|\psi(x)|^2$ for finite $M_z$ with $|\psi(x)|^2$ for a finite $M_x$ of the same strength, we find that $|\psi(x)|^2$ is not as localized at the S/F interfaces for $M_x$, but rather constant in the whole F region, as shown in Figs.~\ref{fig:ABSx}(d-f).

We also depict the expectation values of the spin densities $\psi^\dagger(x)\sigma_x\psi(x)=\langle\sigma_x(x)\rangle$ and $\psi^\dagger(x)\sigma_y\psi(x)=\langle\sigma_y(x)\rangle$ in Fig.~\ref{fig:ABSx}.
The spin densities $\langle\sigma_x(x)\rangle$ and $\langle\sigma_y(x)\rangle$ are related to the charge currents in $x$ and $y$ directions, respectively (see Sec.~\ref{Sec:CO} below).
By comparing right and left columns of Fig.~\ref{fig:ABSx}, we find that for in-plane magnetization $\langle\sigma_y(x)\rangle$ is delocalized within the F region.
This is in contrast to the out-of-plane case, where the $\langle\sigma_y(x)\rangle$ spin density and the wave functions are peaked near the S/F interfaces.
Another important observation is that for finite $M_x$ the spin polarization amplitudes of the two Andreev levels are no longer equal.
Together with the asymmetry of the Andreev spectrum for $M_x\neq0$ discussed above, a finite $\langle\sigma_y(x)\rangle$ such as in Fig.~\ref{fig:ABSx} gives rise to a finite net Josephson Hall current, even for small $M_x$.
The emergence of this Josephson Hall current will be discussed next.

\section{Current operators and continuity equations}\label{Sec:CO}
Having found ABS with a peculiar behavior for $M_x\neq0$, we next study whether this gives characteristic signatures in observable quantities, such as for example the Josephson current.
In order to derive current density operators, we consider the continuity equation for the charge density defined by the operator
\begin{equation}
\hat{\rho}(\vec r)=e\sum_{\sigma}\hat\psi_{\sigma}^{\dagger}(\vec r)\hat\psi_{\sigma}(\vec r)
\end{equation}
or equivalently by the matrix $\frac{1}{2}e\tau_z\sigma_0$ in the Nambu basis with $e$ denoting the electron charge.
The time evolution of the density operator is given by the equation of motion $\partial\hat{\rho}/\partial t=\i\left[\hat H_{\mathrm{BdG}},\hat{\rho}(\vec r,t)\right]$.
After using the fermionic commutation relations for field operators, this equation of motion can be written in the form of the continuity equation
\begin{equation}\label{eq:drho_dt}
\frac{\partial\hat{\rho}}{\partial t} + \nabla\hat{\vec j}(\vec r)=\hat{S}(\vec r).
\end{equation}
Here, the quasiparticle part of the current density is proportional to the spin operator, analogous to the non-superconducting case for Dirac materials~\cite{Scharf2016:PRL}
\begin{equation}\label{eq:CurrentOp}
\hat{\vec j}(\vec r)=\frac{1}{2} e v_F \hat\Psi^{\dagger}(\vec r)\tau_0\vec{\sigma}\hat\Psi(\vec r).
\end{equation}
The source term corresponding to the conversion of quasiparticles to Cooper pairs in the S leads is given by
\begin{equation}\label{eq:SourceOp}
\hat{S}(\vec r)=\Delta(x)\hat\Psi^\dagger(\vec r)\left[\tau_x\sin\Phi(x) + \tau_y\cos\Phi(x)\right]\hat\Psi(\vec r).
\end{equation}
The expectation values of these one-body operators can be expressed as traces of the Green's function which will be derived in the next section.

\section{Green's function analysis}\label{Sec:GF}

\begin{figure*}[t]
  \subfloat[Even frequency singlet]{%
    \centering
    % \frame{%
      \includegraphics[width=8cm,trim=0cm 0cm 0cm 1.5cm,clip]{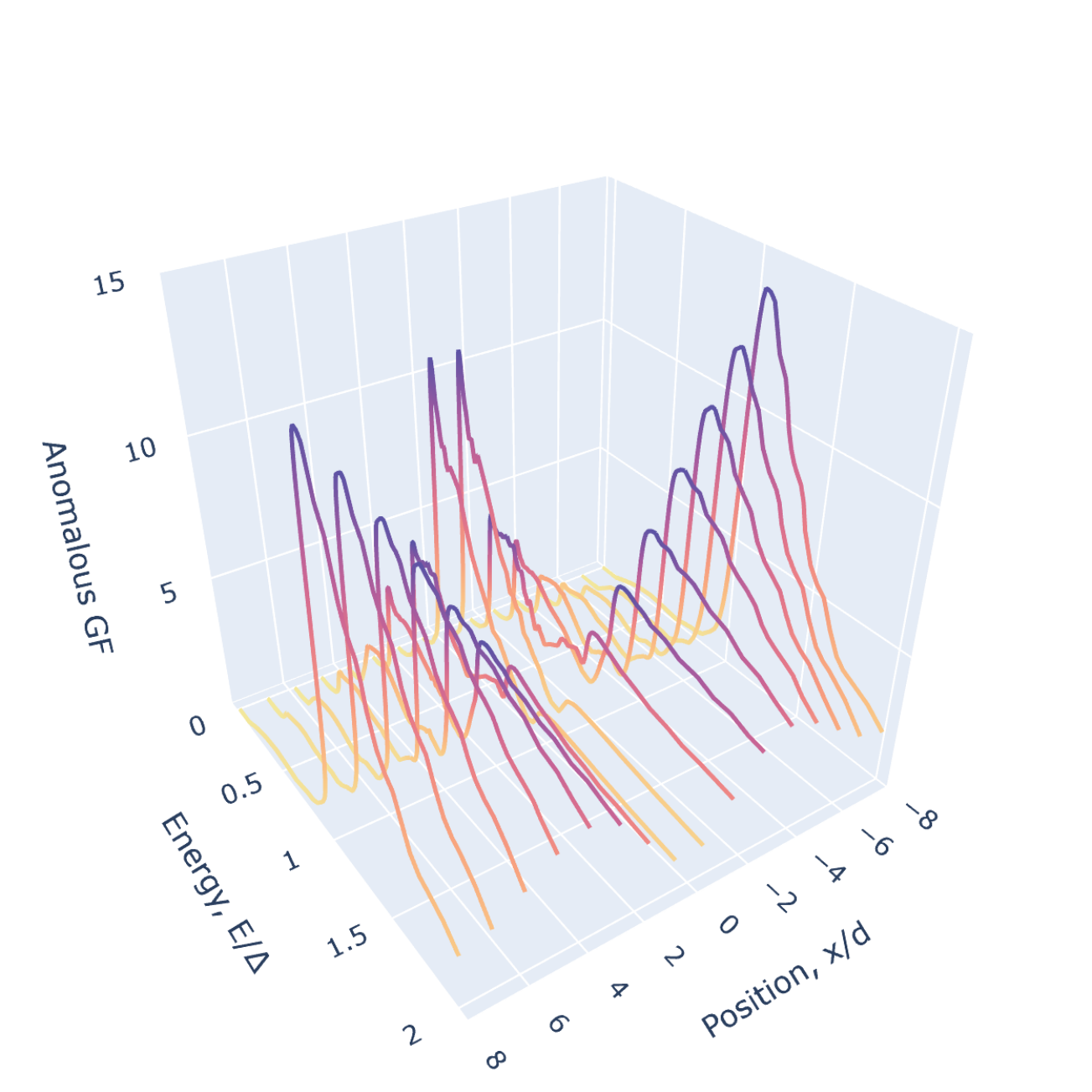}%
    % }
  }
  \subfloat[Odd frequency triplet]{%
    \centering
    % \frame{%
      \includegraphics[width=8cm,trim=0cm 0cm 0cm 1.5cm,clip]{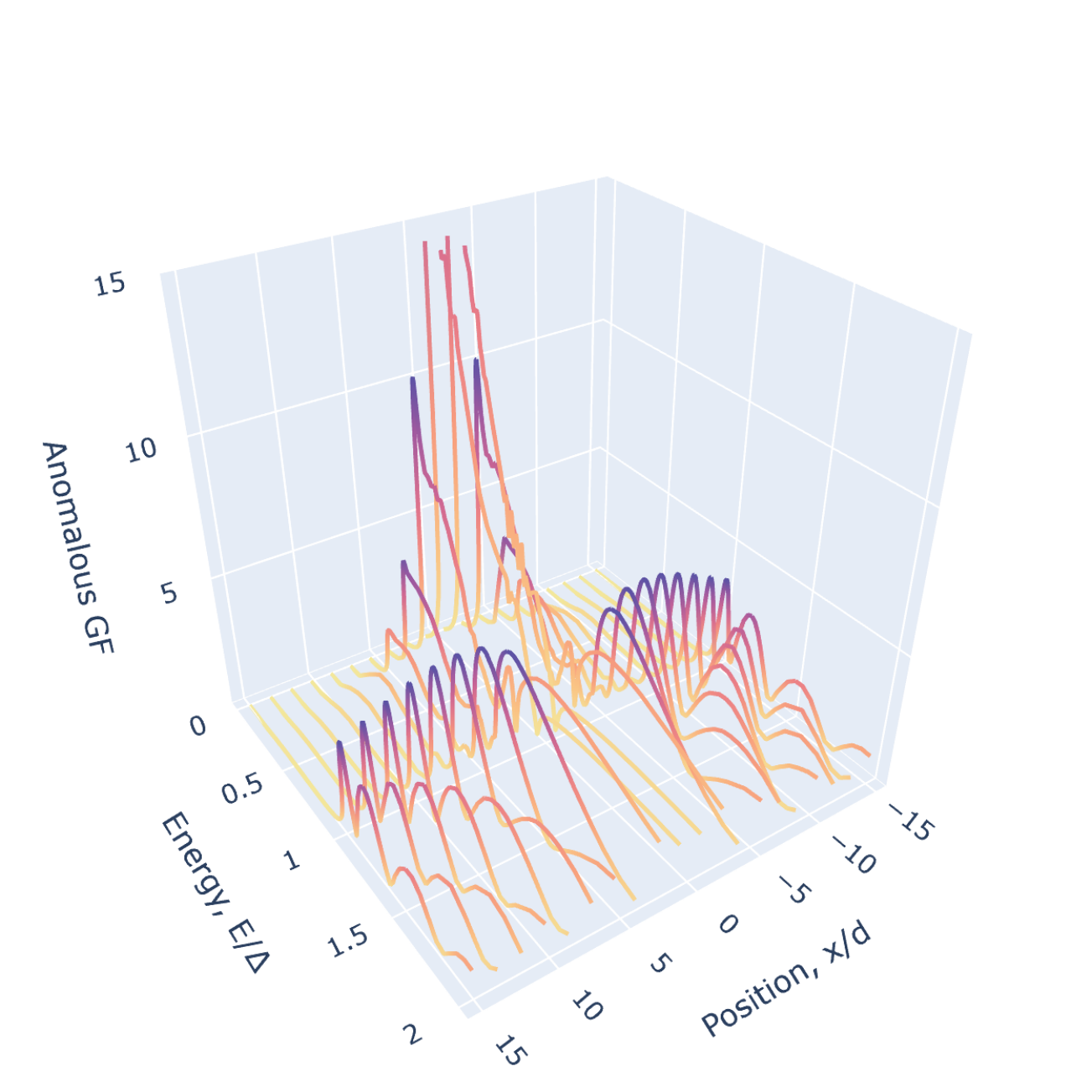}%
    % }
  }
  \caption{
    Position dependent spectrum of anomalous Green's function with magnetization $M_x=0.3$~meV.
    We choose the following parameters:
    $v_F=5\times 10^5$~m/s, $\mu=2$~meV, $d=330$~nm, $\Delta=0.2$~meV, $V_0=0.$~meV, $\phi=0.7\pi$.
    The broadening of the peaks in the spectrum is $\eta=0.02\Delta$.
  }\label{fig:GF-x}
\end{figure*}

\subsection{General expressions and analytical continuation}

In this section, we briefly describe the procedure for constructing the Green's function of the junction Hamiltonian~(\ref{eq:HBdGHamRot}).
We follow the McMillan approach~\cite{McMillan1968} and derive it from the wave-function solutions of the system.
Thus, we choose the energies $\abs{E}>\abs\Delta$, where Eq.~(\ref{eq:PsiLead}) describes propagating states, and solve the scattering problem
\begin{equation}\label{eq:PsiScattering}
\psi_{n_>}(x)=\begin{cases}
  \psi^{(S)}_{n}(x)+\sum_{n'_{<}}r_{nn'}(p_y)\psi^{(S)}_{n'}(x) & x<\frac{d}{2},       \\[5pt]
  \sum_{n'_{>,<}} s_{nn'}(p_y) \psi^{(F)}_{n}(x)                & \abs{x}<\frac{d}{2}, \\[5pt]
  \sum_{n'_{>}}t_{nn'}(p_y)\psi^{(S)}_{n'}(x)                   & x>\frac{d}{2},
\end{cases}
\end{equation}
where the multiindex $n=(\alpha,\xi)\in\{n_{>}\}$ corresponds to an incident state from the left with fixed $p_y$.
Using the boundary conditions defined in Eq.~(\ref{eq:BCfinite}) [or in Eq.~(\ref{eq:BCdelta}) for a $\delta$-barrier], we find the reflection and transmission coefficients $r_{nn'}(p_y)$ and $t_{nn'}(p_y)$ correspondingly.
Analogously, we can obtain states $\psi_{n_<}(x)$ corresponding to processes when there is a quasiparticle incident from the right part of the junction.
These solutions determine the continuum spectrum of the system, located above the superconducting gap.
Furthermore, we employ the same procedure for the transposed Hamiltonian $\hat{H}_{\textrm{BdG}}^T$ to find the conjugate states
\begin{equation}
\hat{H}^T_{\textrm{BdG}} \tilde \psi_n(x) = E\tilde\psi_n(x),
\end{equation}
where the transpose operation acts on the Pauli matrices (Nambu space) and on the coordinate space (by replacing $\hat{\bm{p}}$ with $-\hat{\bm{p}}$).
Afterward, we can write the Green's function for a fixed $p_y$ as an outer product of these solutions
\begin{equation}\label{eq:GFfromPsi}
G^R_{p_y}(x,x',E) = \begin{cases}
  \sum_{n_>,n'_<} C_{nn'} \psi_{n}(x)\tilde\psi_{n'}(x'), & x > x', \\
  \sum_{n_<,n'_>} C_{nn'} \psi_{n}(x)\tilde\psi_{n'}(x'), & x < x',
\end{cases}
\end{equation}
where the position-independent coefficients $C_{nn'}$ should be determined from the boundary condition at $x=x'$,
\begin{equation}\label{eq:GFbc}
G^R_{p_y}(x+0^+,x)-G^R_{p_y}(x-0^+,x)=\frac{\i}{v_F}\tau_z\sigma_x .
\end{equation}

Having determined the Green's function of the system in this way, we can express a given single-particle operator in terms of this Green's function.
In equilibrium the expectation value of the operator can be obtained by evaluating a sum over the fermionic Matsubara frequencies $\omega_n$.
This step requires us to extend the Green's function into the complex plane.
We use the fact that the retarded (advanced) Green's function is analytical in the upper (lower) half of the complex plane.
Hence, we perform an analytical continuation from the open interval on the real axis (given by propagating solutions) to the Matsubara frequencies, by replacing $E\to\i\omega_n$ in all expressions in Eq.~(\ref{eq:GFfromPsi}) \cite{Furusaki1991}.
To access negative Matsubara frequencies, we calculate the advanced Green's function in the same manner as the retarded one.
The uniqueness of the analytical continuation allows us to use these expressions to go to energies below the gap,
so the expectation values obtained in this way contain both contributions from the continuum spectrum and bound states.

Finally, the expectation value of the quasiparticle part of the current density operator is given by
\begin{equation}\label{eq:cws}
j_l(x)\equiv\langle{\hat{j}_l(x)}\rangle=\frac{e v_F}{2\beta}\int dp_y\sum_{n=-\infty}^{\infty}\trace\left[\tau_0\sigma_l G_{p_y}(x,x,\i\omega_n)\right]
\end{equation}
with $l=x,y$~\footnote{Due to translational invariance in the $y$ direction, the currents are constant as a function of $y$ and exhibit only a dependence on the $x$ coordinate.
}.
If $M_x\neq0$, the summation in frequency space for $j_y(x)$ does not converge due to an oscillating behavior at high energies.
This is similar to the behavior of $j_y(x)$ in the normal state, where the contributions arising from the oscillating wave functions for $M_x\neq0$ vanish only after integration over $x$, that is, when computing the transverse current from the transverse current density.
Such a behavior can also be understood as an artifact of the continuum Dirac model.
In fact, this model is only valid close to the Dirac point within the band gap of the TI.
To account for this, we separate the current contributions into superconducting and normal parts, $\vec j=\vec j^{SC}+\vec j^{N}$, where we define $\vec j^{SC}=\vec j-\vec j^{N}$.
Here, $\vec j^N$ is evaluated for a normal system where we set $\Delta=0$ and captures all divergent terms that we treat in more details in Appendix~\ref{App:jnDetails}.
In the remaining expression $\vec j^{SC}$, which is also the part that does not vanish after integration over $x$, the sum converges fast and is performed numerically up to a cutoff.
Since it can be proven that the normal part goes to zero in equilibrium, we focus only on the regular part $\vec j^{SC}$ which describes the actual Josephson current in the junction.

Note that Eq.~(\ref{eq:cws}) only contains the spatial dependence of the quasiparticle part of the current density.
In order to compute the spatial dependence of the full current density, one also needs to include contributions due to the source term $\hat{S}(x)$ from Eq.~(\ref{eq:SourceOp}) in the S leads~\cite{BTK1982:PRB}.
As a consequence the full current density in $x$ direction, consisting of $j_x(x)$ from Eq.~(\ref{eq:cws}) and a term originating from $\hat{S}(x)$ in the S regions, has a constant value and is independent of the position $x$.
For the transverse current, there is no contribution due to $\hat{S}(x)$.
Finally, we remark that the current densities computed from Eqs.~(\ref{eq:GFfromPsi}) and~(\ref{eq:cws}) are the current densities for a situation where all states have equilibrium occupations without any external constraints.
Therefore, Eqs.~(\ref{eq:GFfromPsi}) and~(\ref{eq:cws}) describe the current densities without conservation of the fermion parity.

\subsection{Induced superconducting pairing}

\begin{figure*}
  \centering
  \includegraphics[]{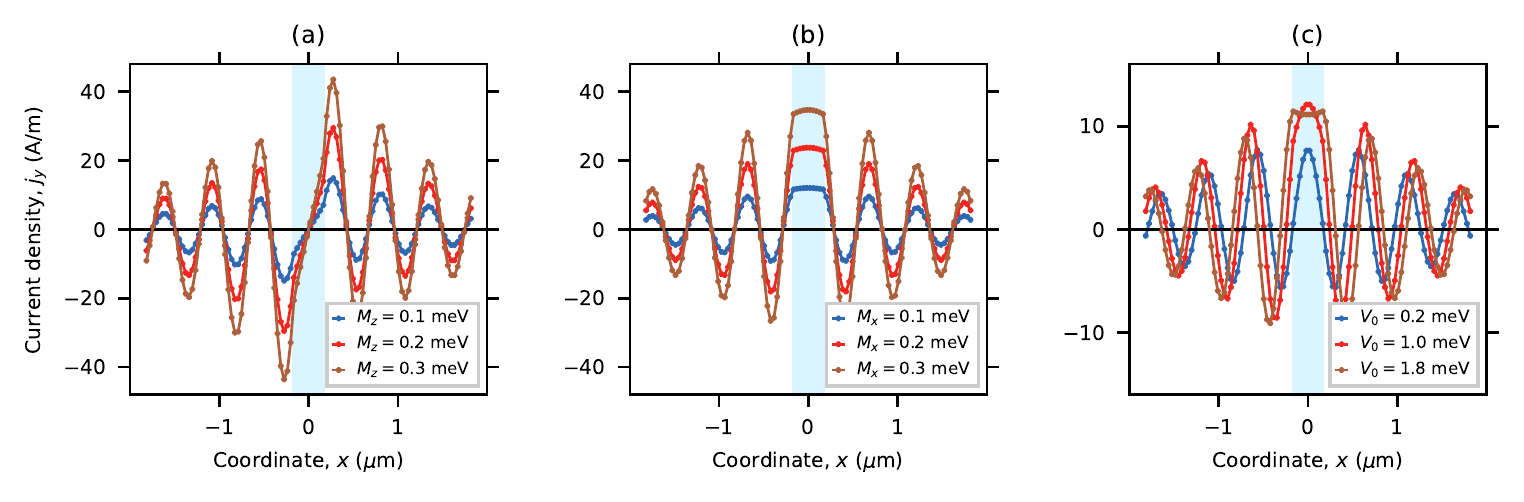}
  \caption{(Color online) Spatial dependence of the transverse current density for different magnetization directions and amplitudes of $\bm{M}$ and $V_0$: (a) $\bm{M}=M_z\bm{e}_z$ and $V_0=1.5$ meV, (b) $\bm{M}=M_x\bm{e}_x$ and $V_0=1.5$ meV, (c) $M_x=0.1$ meV and different $V_0$.
    In all panels, $\phi=0.7\pi$, $d=330$ nm, $\mu=2$ meV, $v_F=5\times10^5$ m/s, and $\Delta=200$ $\mu$eV.
  }\label{fig:JyX}
\end{figure*}

Before discussing the current operators it is interesting to consider how the superconducting correlations are modified in the presence of the S/F/S junction.
The bulk Green's function of a TI based superconductor \cite{Alicea2012:RPP,Tkachov2013:PRB} has mixtures of even frequency singlet $s$-wave and triplet $p$-wave components.
Near the S/F/S interface the translational symmetry is broken and odd-frequency components appear \cite{Tanaka2012:JPSJ,Burset2015:PRB}.
We define the anomalous spectral function as
\begin{equation}
F(x=x',p_y,E) = (F^R - F^A) \tanh\frac{\beta E}{2}.
\end{equation}
Here $F^{R/A}$ are off diagonal parts of the corresponding full Green's function obtained in Eq.~\ref{eq:GFfromPsi}.
The superconducting GF can be further decomposed into singlet and triplet components
\begin{eqnarray}\label{eq:Fdef}
  F_\textrm{SE} (E) & = & \abs{\int dp_y \trace\frac{1}{2}\sigma_0 (F(E) + F(-E))}, \\
  F_\textrm{TO} (E) & = & \sqrt{\sum_{i=1}^3 \abs{ \int dp_y \trace\frac{1}{2}\sigma_i (F(E) - F(-E))}^2 },
\end{eqnarray}
where we have extracted the even- and odd-frequency dependence correspondingly.
Other parts in the GF have odd momentum dependency and hence vanish after $p_y$ integration.
As seen in Fig.~\ref{fig:GF-x}(a, b) the spectrum is composed of the continuum ($E >\Delta$) and bound states ($E<\Delta$) parts.
The continuum spectrum shows the familiar superconducting peak at the gap boundary.
The spectrum inside the SC gap attributed to the bound states is located inside the barrier and decays exponentially away from the barrier.
The even-frequency singlet (SE) part is close to zero inside the gap and recovers full value deep inside the SC lead.
The odd-frequency (TO) part originating from the breaking of the translational symmetry in $x$ direction has its maximum value near the S/F interface and decays into the bulk.
The bound states spectrum is equally composed of SE and TO components.
The TO component can be related to Majorana modes at S/F interfaces \cite{Snelder2015},
which in turn form Andreev bound states in the junction. 
We did not find significant changes in the spectrum of anomalous Green's function between in-plane and out-of-plane magnetization, so we cannot attribute the appearance of the odd-frequency to the finite transverse supercurrent found in the next section.

\section{Josephson Hall current and current-phase relation}\label{Sec:CPR}
We are now in a position to discuss the emergence of the transverse Josephson Hall current, which is the main result of this manuscript.
Without a barrier magnetization, $\bm{M}=\bm{0}$, or if there is only an $M_y$ component of $\bm{M}$, the transverse current density $j_y(x)=\langle\hat{j}_y(x)\rangle$ vanishes.
On the other hand, the asymmetry in the Andreev spectrum due to a finite $M_x$ or the separation of Majorana modes localized at the S/F interfaces due to a finite $M_z$ induce a finite $j_y(x)$.
This is illustrated by Fig.~\ref{fig:JyX}, where we present the spatial dependence of $j_y(x)$ in the presence of a finite magnetization in the barrier.
For a magnetization $M_z$ [Fig.~\ref{fig:JyX}(a)], we observe two transverse current densities of opposite sign localized at the S/F interfaces.
At each interface, this localized current density corresponds mainly to the chiral Majorana mode that emerges at an S/F interface for large $M_z$ as discussed above in Sec.~\ref{Sec:ABSnum}.
The magnitude of $j_y(x)$ increases proportional to $M_z$.
In contrast to the constant longitudinal Josephson current density, $j_y(x)$ oscillates with $k_F$ and decays exponentially into the S regions.
As shown from a symmetry argument in Appendix~\ref{App:jySymm}, $j_y(x)$ is odd with respect $x$ and consequently the total Josephson Hall current through the F region,
\begin{equation}\label{eq:JHallCurrentN}
I_y=\int\limits_{-d/2}^{d/2}\d x\,j_y(x),
\end{equation}
is zero for finite $M_z$.

For a magnetization $M_x$, there is a finite transverse Josephson current density flowing in the same direction inside the whole F region, as shown in Fig.~\ref{fig:JyX}(b).
In this case, the current density profile $j_y(x)$ is an even function of $x$, which clearly allows for a finite Josephson Hall current $I_y$ as given by Eq.~(\ref{eq:JHallCurrentN}) flowing in the F region.
To increase $I_y$, one can apply an additional gate voltage $V_0$ inside the barrier, which reduces the effective Fermi momentum in the F region and hence suppresses the oscillating behavior inside the barrier.
In Fig.~\ref{fig:JyX}(c), one can see that by tuning $V_0$ close to $\mu$ we can achieve an almost flat profile of $j_y(x)$ within the junction, thereby increasing $I_y$.

\begin{figure}[t]
  \centering
  \includegraphics[]{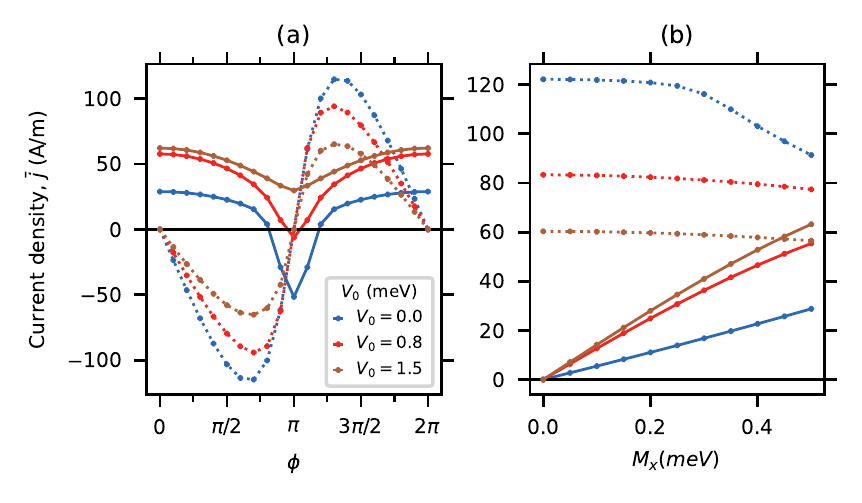}
  \caption{(Color online) Dependence of averaged Josephson current density through the F region (a) on the phase difference $\phi$ for $M_x=0.4$ meV and (b) on the magnetization $M_x$ at $\phi=3\pi/2$ for different gating potentials $V_0$.
    The solid line represents the Josephson Hall current density $\bar{j}_y$ and the dotted line represent longitudinal current density $j_x$.
    In all panels, $d=330$ nm, $\mu=2$ meV, $v_F=5\times10^5$ m/s, and $\Delta=200$ $\mu$eV.
  }\label{fig:JyPhi}
\end{figure}

For the case of $M_x\neq0$ we, moreover, compare the averaged transverse current density $\bar{j}_y = I_y/d$ with the corresponding longitudinal Josephson current density $j_x = I_x/W$ in Fig.~\ref{fig:JyPhi}
\footnote{The longitudinal current density is constant across the junction, so no averaging is needed}.
Here $W$ is the width of the junction, and we normalized the current to the corresponding junction cross-section.
$I_x$ is proportional to the number of transverse modes i.e. $W$, and therefore $j_x$ is effectively $W$ independent.
Figure~\ref{fig:JyPhi}(a) shows the current-phase relation of $j_x$ and $\bar{j}_y$ for several different values of $V_0$.
%\textcolor{red}{[update plot]}
Both $I_x$ and $I_y$ are $2\pi$-periodic in the superconducting phase difference $\phi$ since fermion parity is not conserved if all states have equilibrium occupations (see Sec.~\ref{Sec:GF}).
There is, however, a marked difference in the current-phase relation between the non-sinusoidal $I_x$, which is an odd function of $\phi$, and $I_y$, which is an even function of $\phi$.
Unlike $I_x$, $I_y$ does typically not exhibit zeros at integer multiples of $\phi=\pi$.
Remarkably, we see that for $\phi$ close to $\pi$ the direction of the current can be controlled not only by the sign of $M_x$, but also by modifying the gate voltage $V_0$, which can be appealing for practical applications.

In Fig.~\ref{fig:JyPhi}(b), we show $j_x$ and $\bar{j}_y$ at $\phi=3\pi/2$ as a function of $M_x$.
This illustrates that for a large enough magnetization the Josephson Hall current density can exceed the longitudinal Josephson current.
Such ratios $\bar{j}_y/j_x>1$ are comparable to the ratios found in normal TI-based ferromagnetic tunneling junctions and are a result of the strong SOC in 3D TI surface state.
Although in our theoretical treatment we considered the limit of the system infinite in $y$ direction,
we believe that qualitatively our results also hold for the junctions where $W$ and $d$ are comparable.
This makes the planar Josephson Hall effect in TI-based Josephson junctions a promising candidate to observe sizable transverse currents with ratios $I_y/I_x$ exceeding the corresponding ratios of other Josephson Hall effects~\cite{Yokoyama2015:PRB,Malshukov2019:PRB,Costa2020:PRB}
\footnote{
A similar situation exists for normal tunnel junctions, where the ratio $I_y/I_x$ of the tunneling planar Hall effect~\cite{Scharf2016:PRL},
the normal-state analog to the planar Josephson Hall effect studied here, 
can vastly exceed the corresponding ratio of the tunneling anomalous Hall effect in semiconductors \cite{MatosAbiague2015:PRL},
the normal-state analog to the anomalous Josephson Hall effect in semiconductors~\cite{Costa2020:PRB}.
}.%\cite{MatosAbiague2015:PRL}
We also expect the appearance of transverse Josephson currents (but with smaller $I_y/I_x$ ration) in the Rashba 2DEG ferromagnet junction
based on the study of the normal planar Hall effect in such system \cite{Shen2020:PRB}.

\section{Conclusions}\label{Sec:Conclusions}
In this manuscript, we have studied Josephson junctions realized on three-dimensional topological insulators which are subject to a Zeeman term in the normal topological insulator region.
Most importantly, we have found that the interplay between the spin-momentum locking of the topological insulator surface state, superconductivity and an in-plane Zeeman field in the normal region gives rise to a net transverse Josephson Hall current.
For this Josephson Hall current to emerge, the in-plane Zeeman field has to have a component parallel to the superconducting phase bias direction [see Fig.~\ref{fig:scheme}(d)].
Since the effect is caused by an in-plane Zeeman term, we refer to it as the planar Josephson Hall effect to also distinguish it from other Josephson Hall effects~\cite{Yokoyama2015:PRB,Malshukov2019:PRB,Costa2020:PRB}.

The emergence of the Josephson Hall current is reflected in an asymmetry and 'tilting' of the Andreev spectrum with respect to the transverse momenta $p_y$.
If sufficiently large, this asymmetry even induces a transition in the Andreev spectrum from a regime with gapless, counterpropagating Majorana modes to a regime with unprotected modes that are unidirectional at small $p_y$.
Due to strong spin-orbit coupling, the planar Josephson Hall effect in topological-insulator-based junctions enables sizable Josephson Hall currents, whose amplitudes can be further modulated by electrostatic and/or magnetic control of the normal region.

Until now, we have mainly discussed Zeeman terms induced into the normal topological insulator region by magnetic proximity effects from a nearby ferromagnet, such as in YIG/(Bi,Sb)$_2$Te$_3$~\cite{Jiang2015:NL,*Jiang2016:NC}, EuS/Bi$_2$Se$_3$~\cite{Katmis2016:N} or (Bi,Mn)Te with thin Fe overlayers~\cite{Vobornik2011:NL}.
Since the planar Josephson Hall effect requires in-plane Zeeman terms, an alternative realization could be by applying an in-plane magnetic field along the phase bias direction in the normal region~\footnote{Orbital effects due to a magnetic field, which are not included in our model, would also be minimized if the magnetic field is applied in plane.
  Only out-of-plane magnetic fields are expected to give sizable orbital effects.
}.
Assuming, for example, an in-plane g factor of $g=10$, an in-plane magnetic field of around $B=0.35$ T corresponds to a Zeeman splitting of 0.1 meV~\footnote{For $g=10$ and $B=0.35$ T, the Zeeman splitting is computed as $g\mu_BB/2=0.1$ meV, where $\mu_B$ is the Bohr magneton.
}, which can already yield sizable Josephson Hall currents flowing through the normal region, as illustrated by Fig.~\ref{fig:JyPhi}(b).
Indeed, in Josephson junctions composed of thin-film aluminium and HgTe quantum wells, which can also act as three-dimensional topological insulators~\cite{Bruene2014:PRX}, in-plane magnetic fields of more than 1 T have been achieved~\cite{Hart2014:NP,Hart2017:NP,Ren2019:N}.

It would be interesting to extend our study to systems finite in the $y$ dimension,
where the transverse Josephson current can lead to various phenomena found in systems with SOC.
In this case additional care should be taken to account for currents arising in the superconducting leads, 
by calculating corrections to the spacial phase dependence in a self-consistent manner~\cite{Bergeret2020}.
Effects such as current circulation patterns near the edges
under hard wall boundary conditions~\cite{Bergeret2020},
interplay between longitudinal and transverse phase biases in a crossed junction setup~\cite{Risinggaard2019:PRB},
and control of the phase difference by non-equilibrium current injection~\cite{Bobkova2016} may occur.
Furthermore, here we have focused on transverse \emph{charge} currents in topological Josephson junctions.
For future research on topological Josephson junctions, it might also prove fruitful to study the role of \emph{superspin} Hall currents~\cite{Linder2017:PRB,Risinggaard2019:PRB} and spin polarizations~\cite{Zutic1999:PRB,Zutic2000:PRB,Vezin2020:PRB}, known from semiconductor/superconductor or ferromagnet/superconductor heterostructures.

\acknowledgments
O.M. is grateful to the Chair of Theoretical Physics~4 of the University of W\"urzburg for its hospitality.
B.S. and E.M.H. acknowledge funding by the Deutsche Forschungsgemeinschaft (DFG, German Research Foundation) through SFB 1170, Project-ID 258499086, through Grant No. HA 5893/4-1 within SPP 1666 and through the W\"urzburg-Dresden Cluster of Excellence on Complexity and Topology in Quantum Matter -- \textit{ct.qmat} (EXC 2147, Project-ID 390858490) as well as by the ENB Graduate School on Topological Insulators.

%%%%%%%%%%%%%%%%%%%%%%%%%%%%%%%%%%%%%%%%%%%%%%%%%%%%%%%%%%%%%%%%%%%%%%%%%%%%%%%

\FloatBarrier
\appendix

\section{Andreev bound states in the \texorpdfstring{$\delta$}{delta}-model}\label{App:ABSdelta}

\subsection{Ansatz and boundary conditions}\label{App:ABSbc}
In Sec.~\ref{Sec:ABS} of the main text, we have presented ABS obtained numerically for a finite F region.
Most of the prominent features of short TI-based Josephson junctions are, however, already captured by the model of a $\delta$-like F region with $h(x)=d\delta(x)$, $\Delta(x)=\Delta$, and $\Phi(x)=\phi\Theta(x)$ in Eqs.~(\ref{eq:HBdGHam}) and~(\ref{eq:HBdGHamRot}).
The advantage of this model is that it allows for a transparent analytical treatment of the ABS with relatively compact expressions.

For a $\delta$-junction, the ansatz to obtain the ABS is similar to Eq.~(\ref{eq:ABSansatz}), with the states $\psi(x<0)$ given by the first line of Eq.~(\ref{eq:ABSansatz}) and $\psi(x>0)$ given by the third line of Eq.~(\ref{eq:ABSansatz}).
Now, the coefficients $A_1$, $A_2$, $B_1$, and $B_2$ have to be calculated from the boundary conditions at $x=0$.
This boundary condition can be obtained by integrating Eq.~(\ref{eq:BdG}) from $x=-\eta$ to $x=\eta$ with $\eta\to0^+$.
The corresponding procedure~\cite{MatosAbiague2003:PRA,Sothmann2016:PRB,Scharf2016:PRL} yields
\begin{equation}\label{eq:BCdelta}
\psi(0^+)=\left(\begin{array}{cc}
    \hat{U}_+ & \bm{0}    \\
    \bm{0}    & \hat{U}_- \\
  \end{array}\right)\psi(0^-),
\end{equation}
where
\begin{equation}\label{eq:Mdelta}
\hat{U}_\pm=\e^{\mp\i Z_y}\left(\begin{array}{cc}
    \left[\cos Z\mp\frac{Z_x\sin Z}{Z}\right]  & \i\frac{\sin Z}{Z}\left(\mp Z_z-Z_0\right) \\
    \i\frac{\sin Z}{Z}\left(\pm Z_z-Z_0\right) & \left[\cos Z\pm\frac{Z_x\sin Z}{Z}\right]  \\
  \end{array}\right)
\end{equation}
with $Z_0=V_0d/v_F$, $Z_l=M_l d/v_F$ with $l=x,y,z$, and $Z=\sqrt{Z_0^2-Z_x^2-Z_z^2}$.

\subsection{\texorpdfstring{$\delta$}{delta}-model at \texorpdfstring{$p_y=0$}{py=0}}\label{App:ABSky0}
First, we look at the case of $p_y=0$, where $v_F(\alpha p_{\xi}+ip_{y})/(\mu+\xi\Omega)=\alpha$.
We invoke the boundary condition~(\ref{eq:BCdelta}) on the first and third lines of Eq.~(\ref{eq:ABSansatz}) and require a nontrivial solution of the resulting system of linear equations.
This then yields the two ABS energies $E=\mathcal{P}E_0(\phi)$, where $\mathcal{P}=\pm1$ denotes the two fermion-parity branches and
\begin{equation}\label{eq:Solky0}
E_0(\phi)=\frac{\Delta\cos\left(\phi/2+Z_y\right)}{\sqrt{\cos^2Z+Z_0^2\sin^2Z/Z^2}}.
\end{equation}
The two ABS given by $E=\pm E_0(\phi)$ exhibit a non-degenerate zero-energy crossing at $\phi=\pi$ if $\bm{M}=\bm{0}$~\footnote{Note that in contrast to the main text and to Appendix~\ref{App:ABSAA}, the sign $\pm$ refers to the parity branch, not to the ordering of the energies.
}.
At this crossing, the ground-state fermion parity changes, and the two branches in Eq.~(\ref{eq:Solky0}) have been chosen such that each branch preserves its fermion parity~\cite{FuKane2009:PRB,Ioselevich2011:PRL}.
As such a non-degenerate zero-energy crossing is protected by the fermion parity, it cannot be removed even for finite $\bm{M}\neq\bm{0}$ (see Refs.~\cite{FuKane2009:PRB,Ioselevich2011:PRL} and Appendix~\ref{App:ZEC}).
The crossing can only be shifted, which is what happens for a finite $M_y\neq0$, where $E_0(\phi)=0$ for $\phi=(2n+1)\pi-2Z_y$ with $n\in\mathbb{Z}$.
This protected crossing is a hallmark of the topological Josephson junction and can also be found in models with finite F region.
At $p_y=0$, the main effect of magnetization components $M_{x,z}\neq0$ is thus to reduce the bandwidth of the ABS and detach them from the continuum states.

The finite magnetization $M_y\neq0$ acts as a vector potential in $x$ direction
and can be absorbed in the phase difference by performing the gauge transformation of the BdG Hamiltonian
$U_x = \exp \left[i \tau_z \chi(x) \right]$ where
\begin{equation}
\chi(x) = \begin{cases}
  -Z_y/2,    & x < -d/2,      \\
  M_y x/v_F, & \abs{x} < d/2, \\
  Z_y/2,     & x > d/2.       \\
\end{cases}
\end{equation}
So, the gauge invariant phase difference is given by $\phi-2Z_y$,
which explains the shift of ABS found in Eq.~\ref{eq:Solky0}.

We also remark that the case of $p_y=0$ is equivalent to a Josephson junction based on a single quantum spin Hall edge if $M_x\to M_y$, $M_y\to -M_z$ and $M_z\to M_x$.
With these replacements, Eq.~(\ref{eq:Solky0}) describes the ABS spectrum of such Josephson junctions in the short junction regime~\footnote{Compare to Ref.~\cite{FuKane2009:PRB} discussing Jospehson junctions based on a single quantum spin Hall edge.
}.

\subsection{\texorpdfstring{$\delta$}{delta}-model in Andreev approximation}\label{App:ABSAA}
Another limit that allows for closed analytical solutions is the case of $|\mu|\gg\Delta$, where we can make use of the Andreev approximation.
If we introduce the angle $-\pi/2<\theta<\pi/2$ via $v_F p_y=\mu\sin\theta$, the eigenstates~(\ref{eq:PsiLead}) are simplified within the Andreev approximation in so far that
\begin{eqnarray}\label{eq:DefSAA}
  s_\mu v_F q_\pm\approx\mu\cos\theta\pm\i\frac{\sqrt{\Delta^2-E^2}}{\cos\theta},\nonumber \\
  \dfrac{v_F(\alpha q_{\xi}+ip_{y})}{(\mu+\xi\Omega)}\approx\alpha\e^{\i\alpha\theta}.
\end{eqnarray}

With these approximations, the condition for a nontrivial solution to Eq.~(\ref{eq:BdG}) can be written as
\begin{equation}\label{eq:TransEqAA}
X^2-2A(\theta)X-B(\theta)=0,
\end{equation}
where $X=\sqrt{\Delta^2-E^2}/E$ and
\begin{eqnarray}\label{eq:TransEqAAdef}
  A(\theta)&=&\frac{Z_x\,\sin Z\,\cos Z\,\sin\theta\,\cos\theta}{Z\left[\cos^2\theta\,\cos^2\left(\frac{\phi}{2}+Z_y\right)+\sin^2\theta\,\frac{\left(Z_0^2-Z_x^2\right)\sin^2Z}{Z^2}\right]},\nonumber \\
  B(\theta)&=&\frac{\cos^2\theta\,\left[\sin^2\left(\frac{\phi}{2}+Z_y\right)+\frac{\left(Z_x^2+Z_z^2\right)\sin^2Z}{Z^2}\right]}{\cos^2\theta\,\cos^2\left(\frac{\phi}{2}+Z_y\right)+\sin^2\theta\,\frac{\left(Z_0^2-Z_x^2\right)\sin^2Z}{Z^2}}.
\end{eqnarray}
From the two solutions of Eq.~(\ref{eq:TransEqAA}), $X=A(\theta)\pm\sqrt{A^2(\theta)+B(\theta)}$, one can see that at a fixed angle $\theta$ the two solutions for the energy $E_{\pm}(\phi,\theta)$ do not come as $E_{\pm}(\phi,\theta)=-E_{\mp}(\phi,\theta)$ if $A(\theta)\neq0$.
This is the case for finite $Z_x$ and finite $\theta$.
Instead, the two solutions can be obtained as
\begin{equation}\label{eq:SolAA}
E_\pm(\phi,\theta)=\frac{\sgn\left(A(\theta)\pm\sqrt{A^2(\theta)+B(\theta)}\right)\Delta}{\sqrt{1+\left(A(\theta)\pm\sqrt{A^2(\theta)+B(\theta)}\right)^2}},
\end{equation}
which only satisfies the weaker condition $E_{\pm}(\phi,\theta)=-E_{\mp}(\phi,-\theta)$ originating from the particle-hole symmetry of the formalism.
While such asymmetric ABS could also be obtained within a semiclassical treatment taking into account only phase effects, such a treatment does not capture the exact details of the Andreev spectrum as obtained from a microscopic treatment such as the one provided here.

If $\theta=0$, Eq.~(\ref{eq:SolAA}) reduces simply to $E_\pm(\phi,\theta=0)=\pm|E_0(\phi)|$ with $E_0(\phi)$ given by Eq.~(\ref{eq:Solky0}).
Note that now the sign $\pm$ does not refer to the parity branch, but instead to positive and negative energies.
Another point worth mentioning with regard to Eq.~(\ref{eq:SolAA}) is that for
$Z_x=0$ it reduces to~\cite{Tkachov2013:PRB,Beenakker1992}
\begin{equation}\label{eq:SolAA0}
E_\pm(\phi,\theta)=\pm\Delta\sqrt{1-T(\theta)\left[\sin^2\left(\frac{\phi}{2}+Z_y\right)+\frac{Z_z^2\sin^2Z}{Z^2}\right]},
\end{equation}
where
\begin{equation}\label{eq:TItrans}
T(\theta)=\frac{\cos^2\theta}{\cos^2\theta+\frac{\left(Z_0^2\sin^2\theta+Z_z^2\cos^2\theta\right)\sin^2Z}{Z^2}}
\end{equation}
is the transmission of a normal/ferromagnet/normal junction with $Z_x=0$~\cite{Scharf2016:PRL}.

For $M_x\neq0$, Eq.~(\ref{eq:SolAA}) exhibits several salient features: A finite $M_x\neq0$ introduces not only an asymmetry in the ABS spectrum at finite $p_y$, but can even lead to a situation where the group velocities in $y$ direction, $v_\mathrm{g}\propto\partial E_{\pm}(\phi,\theta)/\partial \theta$, have the same sign for ABS in the vicinity of $p_y=0$ and $\phi\approx\pi-2Z_y$.
At these momenta, the two ABS propagate in the same direction.
This change from nonchiral, counterpropagating ABS to unidirectional ABS propagating in the same direction occurs for $B(\theta)<0$.
Close to $\phi\approx\pi-2Z_y$, $B(\theta)<0$ is satisfied if $|M_x|>|V_0|$.
Hence, if the Zeeman term in the direction of the phase bias $\phi$ exceeds the mismatch between the chemical potentials of the S and F regions, the ABS close to $p_y=0$ and $\phi+2Z_y\approx\pi$ propagate in the same direction in short junctions.

At this point, it is important to remark that the Andreev approximation~(\ref{eq:DefSAA}) breaks down at large transverse momenta, that is, at momenta close to the Fermi momentum $p_F$.
Because of this, Eq.~(\ref{eq:SolAA}) does not describe the ABS for $p_y\approx p_F$ well.
This is also illustrated by Fig.~\ref{fig:ABS}(d), which shows a comparison between Eq.~(\ref{eq:SolAA}) and the results for a finite barrier without any further approximations.
For small $p_y=\mu\sin\theta$, Eq.~(\ref{eq:SolAA}) is in good agreement with the results of the finite barrier.
Equation~(\ref{eq:SolAA}) cannot, however, capture the appearance of zero-energy ABS that occur at large momenta once the modes close to $p_y=0$ become unidirectional.

\section{Effective low-energy model}\label{App:ABSeffM}
The asymmetry of the ABS spectrum as well as the emergence of unidirectional modes for large $M_x$ and small $p_y$ can be understood from the interplay between the effective spin degree of freedom and $M_x$.
To elucidate the origin of these modes, we employ a simple effective low-energy Hamiltonian.
For a $\delta$-like F region, the BdG Hamiltonian~(\ref{eq:HBdGHamRot}) always supports two ABS.
Following the procedure in Ref.~\cite{Fu2008:PRL}, we derive an effective low-energy Hamiltonian describing these two ABS in the vicinity of the protected crossing at $\phi=\pi-2Z_y$ for small $p_y$.

To do so, we first note that the BdG Hamiltonian~(\ref{eq:HBdGHamRot}) can be written as $\hat{H}_\mathrm{BdG}(p_y)=\hat{H}_\mathrm{BdG}(p_y=0)+v_Fp_y\sigma_y\tau_z$, where we treat the term $v_Fp_y\sigma_y\tau_z$ as a perturbation.
Then, we can take the two parity-conserving ABS $|\pm\rangle$ for $p_y=0$ discussed in Sec.~\ref{App:ABSky0} and project the full Hamiltonian~(\ref{eq:HBdGHamRot}) onto these two states.
This procedure yields the effective Hamiltonian
\begin{equation}\label{eq:Heff}
\hat{H}_\mathrm{eff}=E_0(\phi)\tilde{\sigma}_z+v_0p_y\tilde{\sigma}_y+v_yp_y\tilde{\sigma}_0,
\end{equation}
where $E_0(\phi)$ is given by Eq.~(\ref{eq:Solky0}) and originates from $\hat{H}_\mathrm{BdG}(p_y=0)$.
In Eq.~(\ref{eq:Heff}), the two-level system formed by the two ABS at $p_y=0$ is described by the Pauli matrices $\tilde{\sigma}_l$ ($l=x,y,z$) and the corresponding $2\times2$ unit matrix $\tilde{\sigma}_0$.
Moreover, we have introduced the velocities
\begin{equation}\label{eq:Veff0}
v_0=\frac{\Delta\left(\Delta\cos Z+\mu\frac{Z_0\sin Z}{Z}\right)}{\Delta^2+\mu^2}\frac{\sqrt{1+\frac{Z_z^2\sin^2Z}{Z^2}}}{1+\frac{Z_x^2+Z_z^2}{Z^2}\sin^2Z}v_F
\end{equation}
and
\begin{equation}\label{eq:VeffZ}
v_y=\frac{\Delta\left(\Delta\frac{Z_0\sin Z}{Z}-\mu\cos Z\right)}{\Delta^2+\mu^2}\frac{\frac{Z_x\sin Z}{Z}}{1+\frac{Z_x^2+Z_z^2}{Z^2}\sin^2Z}v_F,
\end{equation}
which arise from the matrix elements of the perturbation $v_Fp_y\sigma_y\tau_z$.
Since we are mainly interested in the ABS close to the crossing at $\phi=\pi-2Z_y$, we have approximated the $\phi$-dependent velocities $v_0(\phi)$ and $v_y(\phi)$ by $\phi$-independent velocities $v_0(\phi)\approx v_0(\pi-2Z_y)\equiv v_0$ and $v_y(\phi)\approx v_y(\pi-2Z_y)\equiv v_y$.

The spectrum of Eq.~(\ref{eq:Heff}) is given by $E_\mathrm{eff}^{\pm}(\phi)=v_yp_y\pm\sqrt{E^2_0(\phi)+(v_0p_y)^2}$.
At the crossing point, $E_0(\pi-2Z_y)=0$ and $E_\mathrm{eff}^{\pm}(\pi-2Z_y)=(v_y\pm v_0)p_y$.
If $v_y=0$, that is, if $M_x=0$, the spectrum at $\phi=\pi-2Z_y$ is simply $E_\mathrm{eff}^{\pm}(\pi-2Z_y)=\pm v_0p_y$ and describes two counterpropagating Majorana modes along the $y$ direction, similar to Ref.~\cite{Fu2008:PRL}.
For finite $v_y$, on the other hand, the group velocities $(v_y\pm v_0)$ of the two modes point into the same direction if $|v_y|>|v_0|$.

The appearance of a term $v_yp_y\tilde{\sigma}_0$ in Eq.~(\ref{eq:Heff}) has thus its origin in the unidirectional modes at small $p_y$.
While there is always a finite $v_0$ in TI-based Josephson junctions, $v_y\neq0$ only arises for finite $M_x\neq0$.
This can be understood in the following way: The terms containing $v_0$ and $v_y$ originate from the matrix elements $v_Fp_y \langle\mathcal{P}|\sigma_y\tau_z|\mathcal{P}'\rangle$ with $\mathcal{P},\mathcal{P}'=\pm1$ denoting the two parity branches of $p_y=0$.
If $M_x=0$, the effective spin orientation of the eigenstates $|\pm\rangle$ of $\hat{H}_\mathrm{BdG}(p_y=0)$ lie in the $xz$ plane and thus the expectation values $\langle\pm|\sigma_y\tau_z|\pm\rangle$ vanish and $v_y=0$.
Only off-diagonal matrix elements $\langle\mp|\sigma_y\tau_z|\pm\rangle$ are finite and give rise to $v_0\neq0$.

For finite $M_x\neq0$, however, the effective spin expectation values of $|\pm\rangle$ now also acquire a component in the $y$ direction and $\langle\pm|\sigma_y\tau_z|\pm\rangle\neq0$.
The eigenstates $|\pm\rangle$ satisfy the relation $|\pm\rangle=\hat{\mathcal{K}}|\mp\rangle$, where $\hat{\mathcal{K}}$ denotes complex conjugation.
Because of this property, $\langle+|\sigma_y\tau_z|+\rangle=\langle-|\sigma_y\tau_z|-\rangle$ and consequently the diagonal matrix elements of the perturbation is proportional to $\tilde{\sigma}_0$ (and not to $\tilde{\sigma}_z$ or a linear combination of $\tilde{\sigma}_0$ and $\tilde{\sigma}_z$).
The spectrum of Eq.~(\ref{eq:Heff}) makes it clear that the ABS spectrum for small $p_y$ and close to the protected crossing point $\phi+2Z_y=\pi$ (or, more generally, close to $\phi=(2n+1)\pi-2Z_y$ with $n\in\mathbb{Z}$) can support unidirectional modes around $p_y\approx0$ for finite $M_x$.

\section{Protected zero-energy crossing for \texorpdfstring{$p_y=0$}{py=0}}\label{App:ZEC}
A peculiar feature of the ABS spectrum of a Josephson junction based on a single surface of a 3D TI is its protected zero-energy crossing for $p_y=0$, even in the presence of a Zeeman term in the F region, as discussed in Sec.~\ref{App:ABSky0}.
Following Ref.~\cite{Ioselevich2011:PRL}, we can understand this protection arising from the particle-hole symmetry of the BdG Hamiltonian, which allows one to define a Pfaffian, $\mathrm{Pf}\left[\hat{H}_\mathrm{BdG}(p_y=0)\right]$ for any $\phi$.
The existence of a Pfaffian then implies that two-fold degenerate zero-energy states are generically protected against perturbations as long as particle-hole symmetry is preserved.

For the system studied here, particle-hole symmetry is described by the operator $\hat{\mathcal{C}}=\sigma_y\tau_y\hat{\mathcal{K}}$, where $\hat{\mathcal{K}}$ denotes complex conjugation and $\sigma_y$ and $\tau_y$ are the respective Pauli matrices in spin and electron/hole space.
Any BdG Hamiltonian, including Eq.~(\ref{eq:HBdGHamRot}), anticommutes with $\hat{\mathcal{C}}$,
\begin{equation}\label{eq:PHS}
\left\{\hat{\mathcal{C}},\hat{H}_\mathrm{BdG}\right\}=0.
\end{equation}
If we introduce the momentum quantum number $p_y$, this becomes
\begin{equation}\label{eq:PHSky}
\hat{\mathcal{C}}\hat{H}_\mathrm{BdG}(p_y)\hat{\mathcal{C}}^{-1}=-\hat{H}_\mathrm{BdG}(-p_y).
\end{equation}
Thus, only for $p_y=0$, does particle-hole symmetry imply $\left\{\hat{\mathcal{C}},\hat{H}_\mathrm{BdG}(p_y=0)\right\}=0$, while in general particle-hole symmetry connects states with $p_y$ to states with $-p_y$.

From now on, we focus only on the two ABS $|\pm\rangle$ at $p_y=0$ and with $\bm{M}=\bm{0}$.
For a $\delta$-like F region and $\bm{M}=\bm{0}$, the energies are simply
given by $E=\pm\Delta\cos(\phi/2)$, that is, they possess two-fold degenerate zero-energy states at $\phi=\pi$.
Similar to Sec.~\ref{App:ABSeffM}, the corresponding low-energy Hamiltonian is the $2\times2$ matrix with respect to the ABS $|\pm\rangle$,
\begin{equation}\label{eq:Heff0}
\hat{H}_\mathrm{eff}^0=\left(\begin{array}{cc}
    \Delta\cos(\phi/2) & 0                   \\
    0                  & -\Delta\cos(\phi/2) \\
  \end{array}\right),
\end{equation}
which can in turn be transformed to
\begin{equation}\label{eq:Heff0t}
\hat{\tilde{H}}_\mathrm{eff}^0=\i\left(\begin{array}{cc}
    0                   & \Delta\cos(\phi/2) \\
    -\Delta\cos(\phi/2) & 0                  \\
  \end{array}\right)\equiv\i\hat{A}_\mathrm{eff}^0.
\end{equation}
The Pfaffian of Eq.~(\ref{eq:Heff0}) is then given by $\mathrm{Pf}\left(\hat{H}_\mathrm{eff}^0\right)=\i\mathrm{Pf}\left(\hat{A}_\mathrm{eff}^0\right)=\i\Delta\cos(\phi/2)$ and can be related to the ground-state fermion parity $F_0$ via $(-1)^{F_0}=\mathrm{sgn}\left[\mathrm{Pf}\left(\hat{A}_\mathrm{eff}^0\right)\right]$.
Since $\mathrm{Pf}\left(\hat{H}_\mathrm{eff}^0\right)$ exhibits only a single zero, a perturbation that preserves particle-hole symmetry cannot remove the two zero-energy states, but only shift them to other values of $\phi$~\cite{Ioselevich2011:PRL}.

Now, we are in a position to understand why the crossing at $\phi=\pi$ is protected against finite $\bm{M}$ in the F region.
For finite $\bm{M}$ and $p_y$, we can write the BdG Hamiltonian as
\begin{equation}\label{eq:FullHam}
\hat{H}_\mathrm{BdG}(p_y)=\hat{H}_\mathrm{BdG}(p_y=0)|_{\bm{M}=\bm{0}}+\hat{H}^{'}_{\bm{M}}+\hat{H}^{'}_{p_y}
\end{equation}
with
\begin{equation}\label{eq:FullHamM}
\hat{H}^{'}_{\bm{M}}=-\bm{M}'\cdot\bm{\sigma}\;h(x)
\end{equation}
and
\begin{equation}\label{eq:FullHamky}
\hat{H}^{'}_{p_y}=v_Fp_y\sigma_y\tau_z.
\end{equation}
We remind the reader that because of the rotated spin axes used in Eq.~(\ref{eq:HBdGHamRot}) $\bm{M}'$ in Eq.~(\ref{eq:FullHamM}) is a rotated effective magnetization.
This magnetization $\bm{M}'$ is related to the components of the real magnetization $\bm{M}=(M_x,M_y,M_z)$ induced in the F region via $\bm{M}'=(-M_y,M_x,M_z)$.
In Eq.~(\ref{eq:FullHam}), the additional terms behave differently under $\hat{\mathcal{C}}$: $\left\{\hat{\mathcal{C}},\hat{H}^{'}_{\bm{M}}\right\}=0$ and thus a finite $\bm{M}$ does not remove the zero-energy crossing.
On the other hand, $\left\{\hat{\mathcal{C}},\hat{H}^{'}_{p_y}\right\}\neq0$ and thus a gap is opened at finite $p_y$ because in this case particle-hole symmetry does not protect $\hat{H}^{'}_{p_y}$, but connects $\hat{H}^{'}_{p_y}$ and $\hat{H}^{'}_{-p_y}$ (see above).
While we have employed this analysis to the case of a $\delta$-barrier for illustration, we note that this is valid for all single-energy crossings that are only double degenerate, including the case of a finite barrier also studied in this manuscript~\footnote{In contrast, such zero-energy crossings are not protected for a two-dimensional electron gas without magnetic field because the crossings are four-fold degenerate due to the spin degeneracy.
}.

Hence, as a final remark we note that the analysis from Eqs.~(\ref{eq:Heff0}) and~(\ref{eq:Heff0t}) applies also to the case of finite $\bm{M}$, where $\Delta\cos(\phi/2)$ should simply be replaced by $E_0(\phi)$ from Eq.~(\ref{eq:Solky0}).
This also makes it clear that the ground-state fermion parity $F_0$ given by $(-1)^{F_0}=\mathrm{sgn}\left[E_0(\phi)\right]=\mathrm{sgn}\left[\cos(\phi/2+Z_y)\right]$ for finite $\bm{M}$ is only shifted in its $\phi$ dependence by $Z_y\propto M_y$, but remains unaltered otherwise.

%%%%%%%%%%%%%%%%%%%%%%%%%%%%%%%%%%%%%%%%%%%%%%%%%%%%%%%%%%%%%%%%%%%%%%%%%%%%%%%%

\section{Additional Andreev bound states results}

\subsection{Large out-of-plane Zeeman term}\label{App:ABSlargeMz}

\begin{figure}[t]
  \includegraphics[width=\columnwidth]{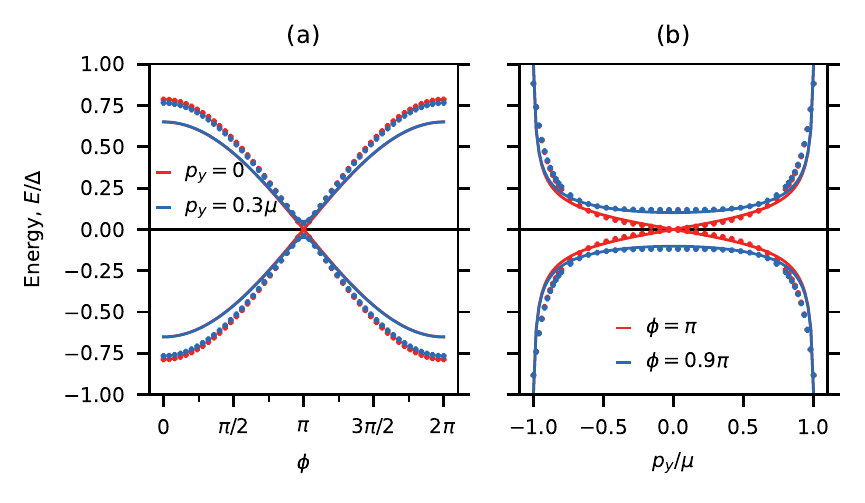}
  \caption{Dependence of the Andreev bound states on (a) phase and (b) transverse momentum. Here, we used the parameters $v_F=5\times 10^5$~m/s, $\mu=2$~meV, $d=330$~nm, $\Delta=0.1$~meV, $V_0=0.2$~meV and $M_z=1$~meV. Solid lines are analytical expressions for the $\delta$-barrier model and dots are obtained from the numerical solution of a finite-barrier problem.}\label{fig:ABS-LargeMz}
\end{figure}

In this part, we provide additional plots of the ABS and discuss the relevant system parameters required to observe the effect of detaching the Andreev bound states from the continuum spectrum.
Inducing a bigger gap between the Andreev bound states and the continuum states requires increasing the effective barrier strength $Z_z$.
This can be achieved either by a stronger magnetic field or magnetization $M_z$ or by increasing the barrier length $d$.
Here we keep $d$ relatively small, i.e. in the short-junction limit, in order to be comparable with our analytical $\delta$-barrier solution.
On the other hand, large $M_z$ requires large g-factors and huge magnetic fields which are not feasible in experiments. If the Zeeman term is proximity-induced from a nearby ferromagnet on the other hand larger $M_z$ are indeed possible.
Figure~\ref{fig:ABS-LargeMz} shows the results for such a case, where we increased the magnetization to $M_z=1$~meV.

\subsection{Chemical potential dependence}\label{App:ABSsmallmu}

\begin{figure}[t]
  \centering
  \includegraphics[]{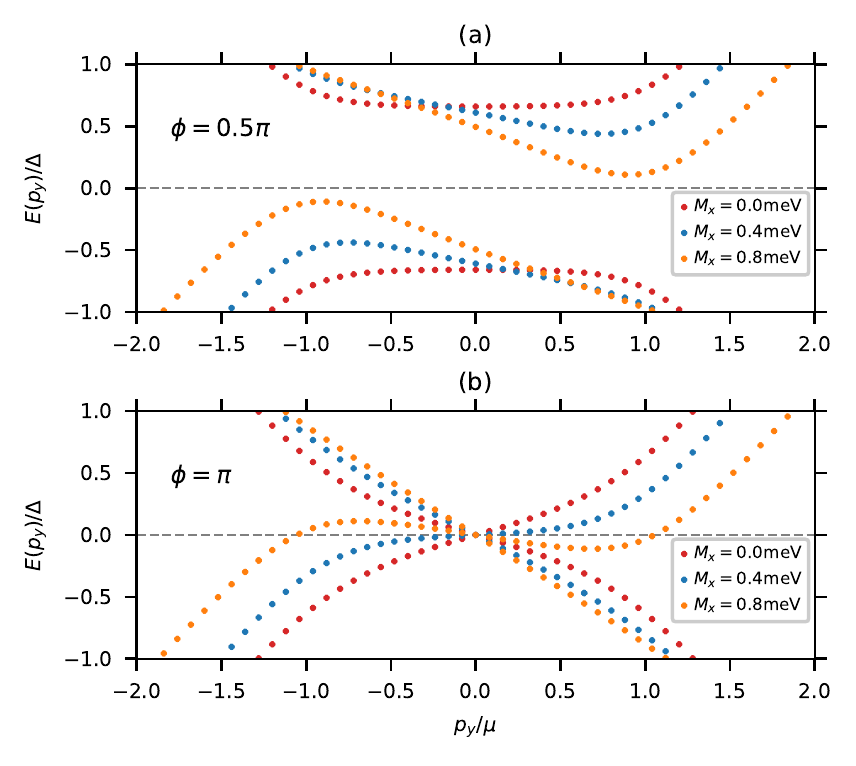}
  \caption{Dependence of the Andreev bound states on the transverse momentum for $\mu=0.2$ meV and different $M_x$ at (a) $\phi=0.5\pi$ and (b) $\phi=\pi$. Here, we use the parameters $v_F=5\times10^5$~m/s, $d=330$~nm, $\Delta=0.1$~meV, $M_y=0$, $M_z=0$ and $V_0=0$.
  }\label{fig:ABS-MxMu0p2}
\end{figure}

Although more difficult to control experimentally,
our model allows also to consider a situation where the chemical potential $\mu$ in the proximitized S leads is similar to the induced superconducting gap, i.e. $\mu \sim \Delta$.
This situation is beyond the Andreev approximation commonly used in previous studies.
We show the Andreev bound states for this regime in Fig.~\ref{fig:ABS-MxMu0p2}.
Similar to the case of large $\mu$, a finite $M_x$ leads to an asymmetry in the Andreev spectrum with respect to the transverse momentum $k_y$,
which in turn gives rise to a transverse Josephson Hall current.
A main effect of the decreased chemical potential is that larger values of $M_x$ are required to achieve the regime with unidirectional modes close to $k_y=0$.
This can be seen by comparing parameters used in Fig.~\ref{fig:ABS-MxMu0p2}(b) and the ones in Fig.~\ref{fig:ABS}(d) in the main text.
Figure~\ref{fig:ABS-MxMu0p2} also illustrates that for small chemical potentials Andreev bound states appear at momenta $k_y$ larger than $k_F$, which is very different from the commonly studied case of $\mu\ll\Delta$.

%%%%%%%%%%%%%%%%%%%%%%%%%%%%%%%%%%%%%%%%%%%%%%%%%%%%%%%%%%%%%%%%%%%%%%%%%%%%%%%%

\section{Normal junction Green's function}\label{App:jnDetails}
Let us consider the simpler case of a normal junction to better understand the asymptotic behavior of the superconducting solution for $\abs E\gg\abs{\Delta}$.
If we switch off superconductivity by putting $\Delta=0$, there must be no current in equilibrium.
However, calculating the expectation value of the transverse current has some technical difficulties which we address in this section.
Without the superconductor, the Hamiltonian is defined by
\begin{equation}\label{eq:HN}
H_N=v_F\vec\sigma \cdot \hat{\vec p}-\mu+(V_{0}-\vec M' \cdot \vec{\sigma})h(x).
\end{equation}
To obtain the Green's function, we proceed analogously to the main text.
For example, the lead solutions are
\begin{equation}\label{eq:wfNLeads}
\psi^{(0)}_{\alpha}(x)=\frac{1}{\sqrt{2}}
\begin{pmatrix}
  1 \\
  v_F\dfrac{\alpha q_0 + ip_{y}}{\mu + E}
\end{pmatrix}e^{i\alpha q_0 x},
\end{equation}
where $v_F q_0=\sqrt{\left(\mu+E\right)^{2}-(v_F p_y)^{2}}$ and $\alpha=\pm 1$ gives the direction of propagation.
In this case, we have two helical counterpropagating sates for each $p_y$.
The F-barrier solution is given by Eq.~(\ref{eq:PsiNormal}) with $\xi=+1$.
We omit the details of solving the transposed Hamiltonian and deriving the scattering states.

To analyze the current expectation value, we stay within the real-energy picture because it allows for a discussion of high-energy contributions and the continuation to the Matsubara frequencies when the function does not decay fast for $|E|\to\infty$.
The current operator simplifies to $\vec j^N = e v_F \vec \sigma$ and we obtain
\begin{equation}
\left\langle j^N_{i}(x)\right\rangle = -2e v_F \int dE \, n(E) %\\
\int dp_{y}\Im\left[\trace\sigma_{i}G^{R}_{p_y}(x,x)\right],
\end{equation}
where $n(E)$ is the Fermi-Dirac distribution.
In the case $\abs{E+\mu}\geq\abs{p_{y}}$, we can conduct a variable substitution in the integral, namely $v_F q_0=\abs{E+\mu}\cos\theta$ and $v_F p_{y}=\abs{E+\mu}\sin\theta$.
This allows us to rewrite the integration limits over $p_{y}$, which yields for the transverse current
\begin{multline}\label{eq:jyNormal}
  \left\langle j^N_{y}(x < -d/2)\right\rangle =2e v_F\int dE\abs{E+\mu}n(E) \\
  \int_{-\pi/2}^{\pi/2}d\theta\Im\left[r(E,\theta)e^{-2ix(\mu+E)\cos\theta/v_F}e^{-i\alpha\theta}\right],
\end{multline}
where $r(E,\theta)$ is the reflection coefficient of the mode incident from the left lead.
The case of $\abs{E+\mu}<\abs{p_{y}}$ is treated analogously employing hyperbolic functions.
In the non-superconducting case, it is possible to obtain a relatively compact form for the reflection coefficient~\cite{Scharf2016:PRL}.

\paragraph{$\delta$-barrier solutions}
We consider the stationary states similar to Eq.~(\ref{eq:PsiScattering}), with the superconducting lead wave functions replaced by $\psi^{(0)}_{n}(x)$ and in the limit $d\to0$.
Then, using the boundary condition~(\ref{eq:BCdelta}) for the electron block, we obtain the reflection and transmission coefficients
\begin{align}
  r & =\frac{e^{is\theta}(Z_{x}+iZ_{z}\cos\theta-sZ_{0}\sin\theta)\sin Z}{Z\cos\theta\cos Z+i(Z_{0}-sZ_{x}\sin\theta)\sin Z}, \\
  t & =\frac{e^{-iZ_{y}}Z\cos\theta}{Z\cos\theta\cos Z+i(Z_{0}-sZ_{x}\sin\theta)\sin Z},
\end{align}
where we have defined $s=\sgn(E+\mu)$.
We notice the property $r(E,\theta)=r(-E,-\theta)$.
The $\delta$-barrier does not introduce an energy scale.
Therefore, the reflection amplitude is energy independent.
This means that all states in the system are affected by the introduction of the barrier, which has significant consequences for Eq.~(\ref{eq:jyNormal}) because the spectrum is not bounded from below.

\paragraph{Finite-barrier solutions}
By using the boundary conditions from Eq.~(\ref{eq:BCfinite}) and lead wave functions defined in Eq.~(\ref{eq:wfNLeads}), we obtain the reflection and transmission coefficients as
\begin{align}
  r(E) = & \frac{e^{is\theta-id(E+\mu)\cos\theta/v_F} a(E)\sin(d k_0)}
  {v_F k_0\cos\theta\cos(d k_0) + ib(E)\sin(d k_0)},
  \\
  t(E) = & \frac{e^{-idM_{y}/v_F-id(E+\mu)\cos\theta/v_F} v_F k_0 \cos\theta}
  {v_F k_0\cos\theta\cos(d k_0) + ib(E)\sin(d k_0)}
\end{align}
with $s=\sgn(E+\mu)$ and
\begin{align}
  a(E) = & M_{x}+iM_{z}\cos\theta-sV_{0}\sin\theta,
  \\
  b(E) = & (E+\mu)\cos^{2}\theta+V_{0}-M_{x}\sin\theta.
\end{align}
In this case, $r(E)$ exhibits a behavior $\sim1/\abs{E}$ for large $\abs{E}$, but this is still not enough to make the energy integral in Eq.~(\ref{eq:jyNormal}) finite.

The divergent behavior comes from the fact that the energy spectrum of the Dirac cone is not bound from below, so formally we have to include all contributions down to $E=-\infty$ in the expectation values.
At the same time, the presence of the magnetic barrier introduces spin polarization into all states, making them contribute to the integral~(\ref{eq:jyNormal}).
In the real system, the low-energy model is invalid for energies far from the Dirac cone located in the gap of the topological insulator.
On the other hand, the high-energy solutions become highly oscillating with wave vector $E/v_F$.
These oscillations cannot be resolved in the real system, which provides another argument why we should drop high-energy terms.
Thus, we choose to use the regularization $e^{-\lambda\abs{E}}$ in the integral.
Then, we can perform the energy integration analytically which yields a prefactor $\lambda$ in front of the expression for the current.
Hence, after taking the limit $\lambda\to0$, $\langle j_y^N \rangle$ vanishes.
When computing the current density in the superconducting case ($\Delta>0$), we subtract $\vec j^N$ expression before performing integration.
After that, the integral can be performed numerically or more conveniently by going to the complex plane and mapping it to the Matsubara sum.

%%%%%%%%%%%%%%%%%%%%%%%%%%%%%%%%%%%%%%%%%%%%%%%%%%%%%%%%%%%%%%%%%%%%%%%%%%%%%%%%

\section{Symmetries of the current operator}\label{App:jySymm}
We can get some insight into the current operator expectation values from a symmetry point of view.
In this section, we provide the conditions for the current density $\langle j_y(x) \rangle$ to be an even or odd function.
First, we note that expressions in the junction Hamiltonian~(\ref{eq:HBdGHamRot}) have the following properties: $\Delta(x)=\Delta(-x)$, $h(x)=h(-x)$, and we can choose $\Phi(x)=-\Phi(-x)$ because only the relative phase is important.
Application of the time-reversal symmetry $\mathcal{T}=i\sigma_{y}\mathcal{K}$ results in the following changes in the Hamiltonian: $\Phi\to-\Phi$ and $\vec M'\to-\vec M'$ (equivalent to $\vec M\to-\vec M$).
Inversion symmetry $\mathcal{I}$ has the effect of $\Phi\to-\Phi$ and $\vec \hat p \to-\vec \hat p$, but does not change the spin.
We also consider two mirror planes $\mathcal{M}_{yz}$ and $\mathcal{M}_{xy}$, which act in the spin space such that $M_{y,z}\to-M_{y,z}$ and $M_{x,y}\to-M_{x,y}$, respectively and both reverse the sign of the kinetic term.
If $M_{x(z)}=0$, we find that $\mathcal{S}_{x(z)}=\mathcal{M}_{yz(xy)}\mathcal{IT}$ is a symmetry of the Hamiltonian.

Next, we derive how the current operator transforms under given symmetries
\begin{equation}
\mathcal{S}_x j_{y}\mathcal{S}_x^{-1}=-j_{y} \; \text{and} \; \mathcal{S}_z j_{y}\mathcal{S}_z^{-1}=j_{y}.
\end{equation}
Using that $S\psi(\vec r)$ can be presented as $U \psi^*(V \vec r)$, where $U$ is a unitary matrix in spinor space and $V$ is an orthogonal transformation in coordinate space, we obtain a relation for the contribution of the operator expectation value from a single state
\begin{equation}
\bra{\mathcal{S}\psi(\vec r)}j_y\ket{\mathcal{S}\psi(\vec r)}=\bra{\psi(V \vec r)}\mathcal{S}j_y\mathcal{S}^{-1}\ket{\psi(V \vec r)},
\end{equation}
where the scalar product is performed only in spinor space.
The expectation value of the total current density is the sum of contributions from all states weighted with the occupation number, which is a function of energy.
If $S$ is the symmetry of $H_\mathrm{BdG}$, states $\ket{\psi_n(x,y)}$ and $\mathcal{S}\ket{\psi_n(x,y)}$ either have the same energy or coincide.
Hence, we obtain
\begin{align}
  \langle j_{y}(x, y) \rangle & = - \langle j_{y}(-x, y) \rangle & \text{if $M_x=0$}, \\
  \langle j_{y}(x, y) \rangle & = \langle j_{y}(-x, -y) \rangle  & \text{if $M_z=0$}.
\end{align}
Since the current is independent of $y$ due to translation invariance, these symmetry relation are generalized to the whole junction.

We also mention that in case of the $\delta$-barrier we may have a discontinuity at $x=0$ and the value of the current would depend on the direction from which we approach the barrier.

%%%%%%%%%%%%%%%%%%%%%%%%%%%%%%%%%%%%%%%%%%%%%%%%%%%%%%%%%%%%%%%%%%%%%%%%%%%%%%%%
\FloatBarrier
\bibliography{BibTopInsAndTopSup}

\end{document}